\documentclass[aps,reprint,pra,superscriptaddress,showkeys]{revtex4-1}
\usepackage{amsfonts, amsmath, amssymb}
\usepackage[pdftex]{graphicx}
\makeatletter
\renewcommand\@biblabel[1]{}
\makeatother
\DeclareMathOperator{\Poiss}{Poiss}
\DeclareMathOperator{\corr}{corr}
\newcommand{\SSS}{{\mathrm{S}}}

\usepackage{titlesec}

\titleformat{\section}
  {\large\bf\center}{\thesection}{1em}{}
\titleformat{\subsection}
  {\bf\raggedright}{\thesubsection}{1em}{}

\begin{document}
\title{Interpreting 16S metagenomic data without clustering to achieve sub-OTU resolution}
\author{Mikhail Tikhonov}
\affiliation{Joseph Henry Laboratories of Physics}
\affiliation{Lewis-Sigler Institute for Integrative Genomics}
\author{Robert W. Leach}
\affiliation{Lewis-Sigler Institute for Integrative Genomics}
\author{Ned S. Wingreen}
\affiliation{Lewis-Sigler Institute for Integrative Genomics}
\affiliation{Department of Molecular Biology, Princeton University, Princeton, NJ 08544, USA}
\date{July 11,~2014}
\begin{abstract}
The standard approach to analyzing 16S tag sequence data, which relies on clustering reads by sequence similarity into Operational Taxonomic Units (OTUs), underexploits the accuracy of modern sequencing technology. We present a clustering-free approach to multi-sample Illumina datasets that can identify independent bacterial subpopulations regardless of the similarity of their 16S tag sequences. Using published data from a longitudinal time-series study of human tongue microbiota, we are able to resolve within standard 97\% similarity OTUs up to 20 distinct subpopulations, all ecologically distinct but with 16S tags differing by as little as 1 nucleotide (99.2\% similarity). A comparative analysis of oral communities of two cohabiting individuals reveals that most such subpopulations are shared between the two communities at 100\% sequence identity, and that dynamical similarity between subpopulations in one host is strongly predictive of dynamical similarity between the same subpopulations in the other host. Our method can also be applied to samples collected in cross-sectional studies and can be used with the 454 sequencing platform. We discuss how the sub-OTU resolution of our approach can provide new insight into factors shaping community assembly.
\end{abstract}
\keywords{16S sequencing / data analysis / oral microbiota / OTU clustering}
\maketitle

\begingroup
\renewcommand{\addcontentsline}[3]{}

\section{Introduction}
Host-associated microbial communities are known to be of tremendous importance for host fitness, improving nutrient uptake, training the immune system, and resisting invasion by pathogens (see, for example, Brestoff \& Artis,~2013; Kamada et al.,~2013; Fredricks (ed.),~2013). Our understanding of these communities, however, remains remarkably poor. The origin, maintenance, and importance of community diversity (Fierer et al.,~2011), the factors determining community stability and resilience (Shade et al.,~2012), and the mechanisms of community assembly (Costello et al.,~2012) are only some of the questions driving this rapidly expanding field.

Although most microorganisms cannot be cultured in a laboratory setting, advances in genome-sequencing technology now allow organisms to be probed in their natural environments. In particular, the 16S rRNA tag-sequencing approach identifies community members using fragments of DNA from the hypervariable regions of the ribosomal 16S gene. The development of this technique and the decreasing cost of high-throughput sequencing have prompted a large number of tag-sequencing experiments, including such large-scale efforts as the Human Microbiome Project or the Earth Microbiome Project. The amount of collected data is growing exponentially. However, our ability to interpret this data still has important limitations.

The \emph{de facto} standard approach to 16S data analysis begins by clustering reads by sequence similarity into ``Operational Taxonomic Units'' (OTUs); see Fig.~1A (Quince et al.,~2009; Kunin et al.,~2010; Huse et al.,~2010). A variety of clustering techniques have been developed and are widely used in popular software tools or packages (Hunt et al.,~2008; Schloss et al.,~2009; Edgar,~2010; Huang et al.,~2010; Edgar et al.,~2011; Quince et al.,~2011; Schloss et al.,~2011; Sul et al.,~2011; Caporaso et al.,~2012; Zheng et al.,~2012; Morgan et al.,~2013; Youngblut et al.,~2013). Despite significant progress in the development of such software, all clustering-based approaches suffer from a major shortcoming (Prosser et al.,~2007; Hamady \& Knight,~2009; Schloss \& Westcott,~2011). Although an OTU is a useful concept for coarse-graining sequencing data, its definition is not biologically motivated, but as its name acknowledges is purely operational. Sequences assigned to a particular OTU are generally presumed to be close phylogenetic relatives and therefore likely to derive from ecologically similar bacterial subpopulations. However, the assumption that 16S sequence similarity is a good proxy for ecological similarity is notoriously problematic (Prosser et al.,~2007; Preheim et al.,~2013). Moreover, OTU assignments are not definitive but depend on both the clustering algorithm and the random seed chosen (Schloss \& Westcott,~2011).

\begin{figure*}[t!]
\includegraphics[width = 0.8 \textwidth]{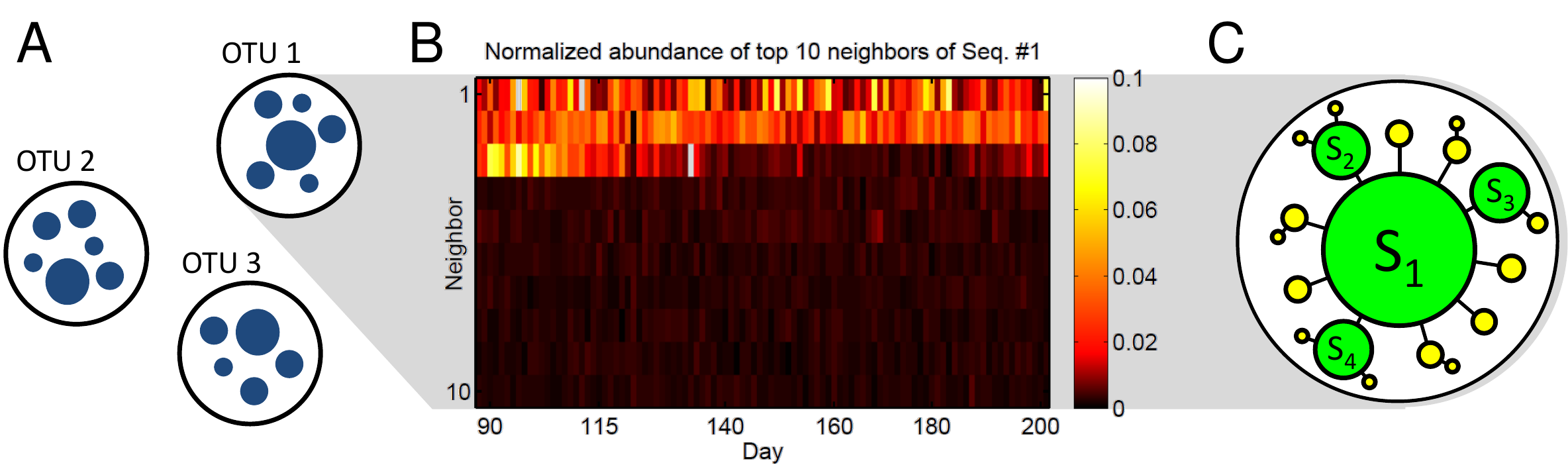}
\caption{\textbf{Clustering reads by OTUs underexploits the quality of modern sequence data.}
\textbf{A.} Cartoon illustrating OTU-based noise filtering. Due to sequencing errors, PCR errors or natural intra-strain variability, each bacterial ``species'' generates a cloud of similar 16S sequences (blue circles; the radius of a circle represents the abundance of a given 16S sequence in a sample, and spacing represents distance in sequence space). Clustering reads into OTUs by sequence similarity is a standard approach to filter this noise. \textbf{B.} Heat map of the abundance, for 100 consecutive samples, of the 10 highest-abundance direct neighbors (Hamming distance = 1) of Seq.~\#1, normalized for each sample to the abundance of Seq.~\#1 (4600 counts/day on average). Three specific direct neighbors are strongly and consistently overrepresented and exhibit distinct dynamics. \textbf{C.} Cartoon based on (B) of the expected structure of an ``error cloud''. Each circle is a unique sequence, with size representing abundance in a sample. True biological sequences ($S_1$--$S_4$; green circles) generate ``daughter'' variants due to substitution errors (yellow circles). Black lines denote $\text{Hamming distance} = 1$ in sequence space.}
\end{figure*}

Several approaches have been proposed to improve the resolution of 16S data analysis beyond the standard 97\%-similarity OTUs. Denoising algorithms exploit the predictable structure of certain error types to attempt to reassign or eliminate noisy reads (Huse et al.,~2010; Quince et al.,~2011; Rosen et al.,~2012). These algorithms are widely used for identifying low-abundance (``rare'') species against a noisy background, often with the aim of improving estimates of ecological diversity. These objectives, however, remain very challenging due to issues that no denoiser can fully address. Any error model is necessarily approximate, and no denoising algorithm can deal with errors that are not adequately described by its error model; when calling low-abundance species this issue becomes particularly problematic. An alternative approach termed Distribution-Based Clustering (DBC; Preheim et al.,~2013) aims to circumvent the limitations of conventional denoisers by using cross-sample comparisons, i.e.\ supplementing sequence information by ecological information (distribution of abundance across multiple biological samples). However, DBC as an OTU clustering algorithm also has important limitations: for low-count sequences, cross-sample comparisons necessarily become unreliable, and the execution time is prohibitively long even for moderately-sized datasets.

Here, we build on the above methods to address a distinct question. Rather than trying to further improve the existing approaches to OTU clustering and rare species identification, we combine error-model based denoising and systematic cross-sample comparisons to resolve the fine (sub-OTU) structure of moderate-to-high abundance community members in 16S Illumina data. Importantly, our method does not rely on clustering similar sequences together. In this regard, our method is similar to oligotyping (Eren et al.,~2013), but our approach does not require manual supervision and applies to an entire community rather than an isolated OTU. Using published data from a longitudinal study where the tongue community of two human individuals was sampled almost daily for several months (Caporaso et al.,~2011), we demonstrate that sequence similarity is a very poor predictor of ecological similarity, which we quantify for two bacteria as the correlation of their abundance time traces (``dynamical similarity''). Thus, most clustering-based approaches would erroneously group together bacterial subpopulations of high ecological diversity for this data set. However, a comparative analysis of the tongue communities of the two individuals also shows that when a pair of 16S tags is observed in both individuals, the dynamical similarity of the pair as measured independently in the two individuals is highly correlated. This correlation falls off substantially when sequences differing by 1 nt out of 130 are compared. In other words, the exact sequence of the 16S tag carried by a bacterial subpopulation is predictive of its ecology, while even 99.2\% similarity between tags of different subpopulations is generally not predictive of dynamical similarity, as defined above. Our results lend support to the recent idea that even a purely 16S-based study can provide insight into functional relatedness of community members (cf. PiCRUST, Langille et al.,~2013), while also exhibiting and beginning to quantify the limitations of such methods. We demonstrate the applicability of our approach to a broad range of dataset types (host-associated longitudinal; environmental cross-sectional; mock community), providing examples when highly similar sequences were found to exhibit ecologically significant distinctions. Finally, we discuss how the single-nucleotide sub-OTU resolution of our method can provide new insights into factors shaping community assembly.

\section{Materials and Methods}
\subsection{Data selection and quality filtering}
We used the raw data from a published long-term longitudinal sampling from four body sites (gut: feces, right and left palm, and tongue) of one male and one female individual (Caporaso et al.,~2011). In this study, the hypervariable region V4 of the bacterial 16S rRNA gene was amplified and sequenced with Illumina GA-IIx. For details on collection and sequencing see the original reference (Caporaso et al.,~2011). Quality-filtered data published with that study is available at MG-RAST:4457768.3-4459735.3 and is sufficient to reproduce our results using provided analysis scripts (see Supplementary Information (SI)). However, to investigate the performance of our filtering approach at different quality filtering settings, for this work we used the demultiplexed, but not quality-filtered FastQ data, kindly provided to us by the study authors. We split this data into per-sample FastQ files using a custom MatLab script (Mathworks, Inc.) and subjected it to minimal quality filtering using USEARCH v.7.0.1090 (Edgar,~2010), truncating reads at Phred quality score 2 (other thresholds were also evaluated; see Fig.~S4), trimming to a fixed length of 130 nt and eliminating reads with ambiguous characters (N). In addition, we removed reads with expected number of base call errors exceeding 1 (maxEE parameter in USEARCH). This criterion only eliminated 1\% of trimmed reads. Notably, our approach does not rely on assumptions about a maximum number of errors in a read. Finally, to facilitate cross-sample comparisons, we compiled a library of all 1.4M unique reads ever observed and a global table listing the abundances of each sequence across samples. This was done using dereplication capabilities of USEARCH and a custom Perl script (\textbf{mergeSeqs.pl}). This script and others referenced in \textbf{bold} below are freely available at \url{https://github.com/hepcat72/CFF}. Finally, the abundance table was normalized to $2.4\,10^4$ total reads per sample, to correct for varying sample size.

Read quality varied across lanes, so the number of reads after quality filtering was highest in a subset of tongue and fecal samples. In this work, we focused primarily on the tongue samples, as these come closest to probing the internal dynamics of a community living in a well-defined location on the body; however, the analysis of fecal samples supports the same conclusions and is presented in Fig.~S11.

Tongue samples were distributed over two lanes. The lane 6 samples from the male subject from day 65 onwards (314 consecutive samples covering a period of 355 days, $2.4 \pm 0.4\,10^4$ reads in quality-filtered samples before normalization) had approximately 4-fold more reads than those from the female subject and from days 1-64 of the male subject (all on lane 5). Consequently, the analysis below uses the data from the male subject from day 65 onwards, and, for the comparative analysis of the two individuals, also the 135 samples collected from the female subject. The early samples from the male subject (days 1-64) are only used for illustrative purposes (Fig.~3D).

To demonstrate the broad applicability of our method we also employed other published data (Figs.~S7 and S11); the data is described in the corresponding legends.

\subsection{Cluster-free filtering}
Clustering can be a useful strategy to coarse-grain 16S data while also reducing noise, but if sequencing noise is low enough, such coarse-graining may not be necessary. At low noise, each community member is predominantly represented by the same 16S sequence, surrounded by a cloud of low-abundance error sequences with the structure of the cloud determined by reproducible error rates. Prior work has described such error clouds in the data (Quince et al.,~2009; Edgar,~2013), and the assumption that high-abundance sequences are likely to be error-free is used in several rank-based denoising and chimera-checking algorithms (SLP, Perseus, Uchime de novo, Uparse, AbundantOTU).

The treatment of reads that are very similar to high-abundance sequences is different across existing algorithms. For example, SLP (Huse et al.,~2010) would consider any read differing by a single nucleotide from a higher-abundance sequence (its ``direct neighbor'' in sequence space) as an error. However, some of these reads may actually represent true community members (Preheim et al.,~2013). A more nuanced treatment can accept a sequence as likely to be real if its observed abundance is highly unlikely to have arisen in error, given some assumptions about error rates. This idea is at the foundation of error-model based denoising. It was used in AmpliconNoise (Quince et al.,~2011), and its recent implementation in DADA (Rosen et al.,~2012) makes DADA, to our knowledge, the best denoiser currently available.

However, no error model is perfect, and for all denoisers, errors not explicitly described by their model are labeled as true sequences. Thus a denoising algorithm alone is insufficient for achieving sub-OTU resolution: if two close sequences that would fall within a single OTU are both identified as ``probably real'', one of these could still be an error. In the context of a single sample, confidently resolving close sequences as independently real requires a different experimental technique (Faith et al.,~2013) or a complete, high-quality reference database of all bacteria in the sample, which in practice is available only for mock communities.

It is possible to resolve this problem in the framework of standard 16S experiments through a comparison of multiple samples, either longitudinal or cross-sectional (Preheim et al.,~2013). As an example, Fig.~1B shows the abundances of the 10 highest-abundance direct neighbors of the overall top sequence of the tongue community, Seq.~\#1, for a representative set of 100 consecutive samples. We see that three specific direct neighbors are strongly and consistently overrepresented compared to the other neighboring sequences and, more importantly, exhibit a dynamical behavior of their own (consider, for example, the 3rd most abundant neighbor). This has a clear interpretation (Fig.~1C): these three sequences must belong to other, fairly abundant bacterial subpopulations, possibly related to Seq.~\#1, but distinct and with their own dynamics.

To achieve sub-OTU resolution, we adopt precisely this strategy, namely a cross-sample correlation analysis of individually denoised samples. Which denoiser should we use? DADA would be an excellent option; however, its estimated execution time on the tongue dataset used here is $2.3\,10^5$ sec (see SI). This is largely due to its exact treatment of probabilities, critically important for the processing of sequences with an abundance of just a few counts. However, for such sequences the imperfections of the error model become non-negligible and cannot be controlled, since cross-sample comparisons are interpretable only for sequences with sufficient abundance. We therefore designed a new, simplified denoiser. Our algorithm, described below, takes two orders of magnitude less time to execute, yet for sequences of moderate abundance considered here achieves performance equal to DADA, as demonstrated using mock community data (Sup. Table S2).

\subsection{Cluster-free filtering -- the denoiser}
For 16S data obtained using the Illumina platform, the main sources of errors are PCR substitutions, PCR chimeras, and substitution errors due to Illumina base call errors. Of these, the substitution errors are responsible for generating the largest number of unique sequences (Fig.~S2; see also Edgar,~2013) and have the most predictable structure: their rates can be estimated directly from the data. To do so, we considered the error clouds around the top 10 sequences by overall abundance (in all tongue samples combined). Assuming that most of these sequences are in fact errors, we determined the rates of specific one-nucleotide substitutions (\textbf{errorRates.pl} with z-score threshold of 2; see SI). These inferred rates were consistent across error clouds observed in the data (Fig.~S3), with the average error rate of only 0.10\% per nucleotide (Sup. Table 1; compare with Quince et al.,~2011, Table 2). We then used these error rates to predict the expected abundance of any given sequence if its presence were entirely due to independently generated sequencing errors of its more abundant neighbors (the ``null model''; Fig.~S5; \textbf{nZeros.pl}). Sequences whose abundance exceeded a threshold of 10 counts and the null-model prediction by at least 10-fold (very conservative filtering parameters), were marked as ``candidates''; their presence cannot be explained as an error within a substitution-only error model (\textbf{getCandidates.pl}). Candidate sequences include true biological 16S sequences, but also sequences that arose through a different type of error, most notably PCR chimeras. Chimeric sequences were identified using UCHIME denovo (Edgar,~2011) on the pooled data from all samples. Most such sequences were already eliminated by the abundance threshold requirement: if we relax the abundance threshold to 2 (excluding singletons only), we find that the chimeras detected by UCHIME, when present in a sample, have abundance under 10 counts in 95\% of cases. However, chimeras of highly abundant parents reproducibly occur at higher abundances (Haas et al,~2011) and are filtered at this step.

Candidate sequences that remained after filtering chimeras were labeled ``real''. Our highly conservative filtering criteria allow us to assume that this list contains only true biological sequences, i.e.\ that there are no false positives (cf. Sup. Table S2), except possibly those due to some exceptionally frequent errors not described by our error model (see SI). This stringency comes at the expense of low-abundance false negatives (true biological sequences labeled as ``possible noise''). Our strategy is to retain all sequences marked ``real'' in 2 or more samples (out of 507; \textbf{getReals.pl}). This makes our denoiser specifically adapted to multi-sample analysis: in each sample, only high-confidence detections are identified, which is very fast, and then a liberal criterion applied across samples retains all sequences that ever generated a high-confidence detection, except sample singletons. In particular, we stress that our detection threshold of 10 counts is not equivalent to removing all sequences with abundance below 10; the only sequences excluded from consideration are those than never rise to 10 counts in the entire set of 504 tongue samples, or do so only once. For such sequences, the measured counts are dominated by detection and counting noise.

In the interest of speed, and to ensure the robustness of reported sequence-abundance values with respect to the details of the error model, we did not attempt to remap noisy reads to their most probable source. Our approach relies on the accuracy of measurement of relative abundances of true sequences. The error remapping process modifies sequence counts in a way that depends on the assumptions of the error model, distorting the relative abundance values whenever neighboring sequences are incorrectly classified as ``reals'' or ``errors''. In contrast, discarding noisy reads leaves the relative abundances intact, as long as the probability of making zero errors is approximately constant across all sequences. This assumption is much weaker than adopting a particular error model. We estimate the zero-error probability at $\approx85$\% (see SI); in other words, discarding noisy reads leads only to a $\approx15$\% loss of sequencing depth. If read remapping is desired, the analysis described below can be applied to DADA denoiser output.

Since non-identical reads are never clustered together, ours is a single-nucleotide resolution approach. The complete workflow of cluster-free filtering is outlined in Fig.~S6 and detailed in the SI. The code is freely available at \url{https://github.com/hepcat72/CFF}.

\begin{figure*}[t!]
\includegraphics[width = 0.8 \textwidth]{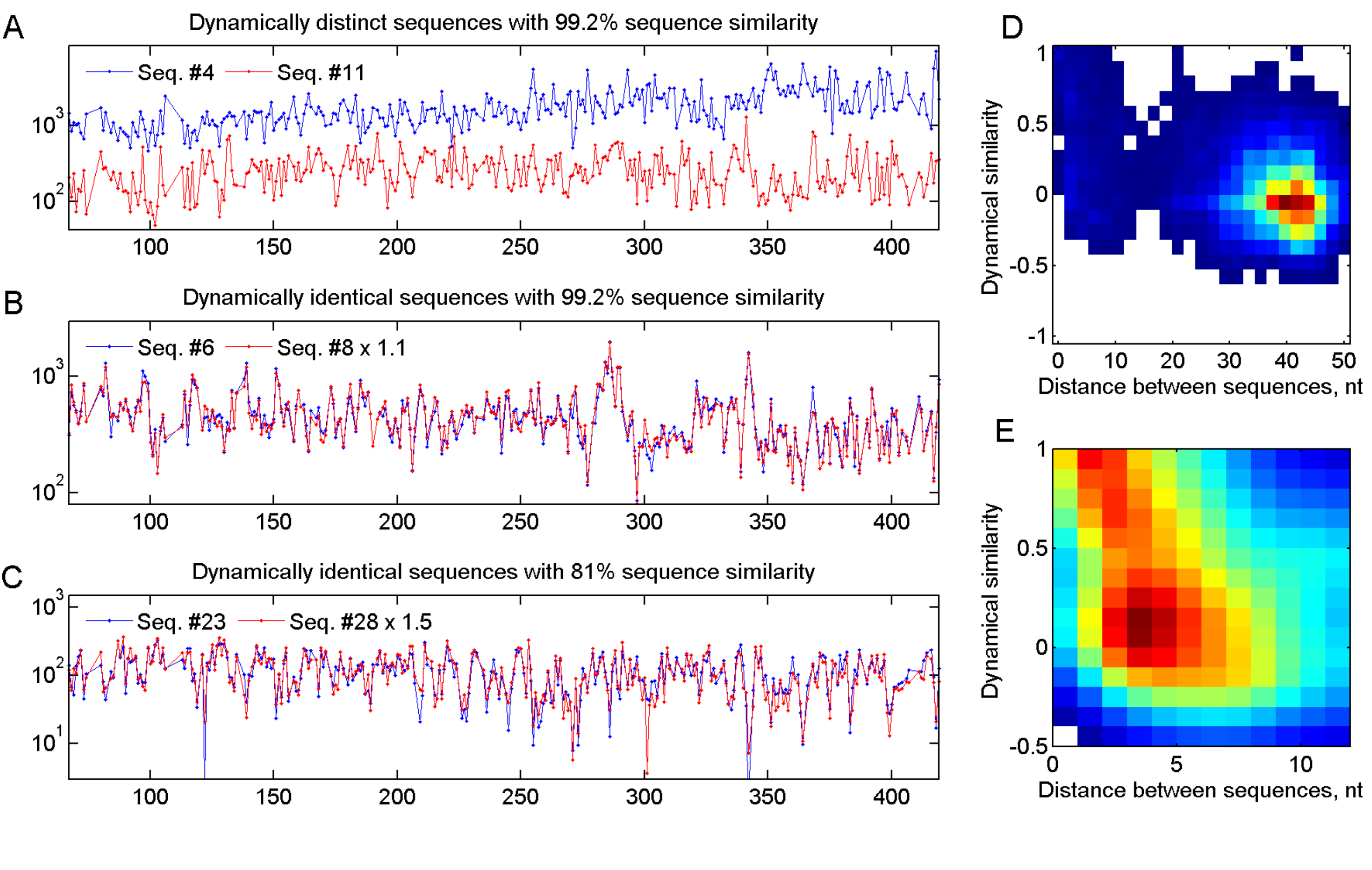}
\caption{\textbf{Sequence similarity need not imply dynamical similarity, and vice versa.}
Panels show sequence counts versus observation day, for days 65-420. \textbf{A.} Seq.~\#4 and \#11, despite 99.2\% sequence similarity, display significant differences in time dependence, indicating that these 16S tags belong to ecologically distinct bacterial subpopulations. \textbf{B.} For Seq.~\#6 and \#8, 99.2\% sequence similarity (1nt difference) is mirrored by near perfect correlation of time series. Red trace renormalized for best overlap. \textbf{C.} Seq.~\#23 and \#28, with only 81\% sequence similarity, nevertheless display near perfect correlation. Red trace renormalized for best overlap. \textbf{D.} 2D histogram of dynamical similarity (Pearson correlation of abundance traces, normalized by maximum expected correlation $c_{\mathrm{max}}$, see text) versus distance in sequence space (nt), for all pairs of the top 200 sequences (19900 data points). \textbf{E.} Zoom-in of D (1321 sequence pairs), showing the most similar sequences. Histogram smoothed for clarity.}
\end{figure*}

\section{Results}
The starting point for our analysis is a global sequence abundance table listing the abundances of each unique 16S sequence across samples. We retained the 307 sequences that passed the multi-sample filtering algorithm described in Methods, and thus putatively belong to bacteria present in the population at least part of the time. We denote these sequences by their overall abundance rank: Seq.~\#1,~\#2, etc. In this list, 184 pairs of sequences were direct neighbors in sequence space (Hamming distance 1). These pairs had 99.2\% sequence similarity but were resolved by our criteria as independently present in the community. The population of bacteria sharing the exact same sequenced fragment of the 16S gene (at 100\% identity) is the smallest taxonomic unit resolvable by 16S analysis. For notational convenience, throughout this work we call it the ``subpopulation'' identified by a sequence.

\subsection{Sequence similarity need not imply ecological similarity, and vice versa}

In the standard approach to tag-sequencing data, it is assumed that sequence similarity of 16S hypervariable regions can be used as a proxy for phylogenetic, and therefore ecological relatedness. Our new filtering method, applied to time-series data, allows us to bypass this assumption and assess ecological relatedness independently, based on the similarity of time traces, since each distinct subpopulation will respond in its own way to variation in environmental conditions (Youngblut et al.,~ 2013), causing the abundance time traces to be more or less correlated (or possibly anticorrelated; see Fig.~S8). Fig.~2A-C illustrates this by showing time traces (normalized counts versus observation day) for three examples of sequence pairs. We find that sequences differing by as little as 1 nucleotide (99.2\% similarity) can be ecologically distinct as evidenced by their very different time series (Fig.~2A); see also Vandewalle et al.,~2012. For comparison, Fig.~2B shows another pair of sequences, also with 99.2\% sequence similarity but whose abundance time traces appear indistinguishable. The remarkable correlation between these two traces provides an internal control and demonstrates that the much lower correlation of traces in Fig.~2A cannot be explained by measurement error but reflects a true ecological difference. Note that the abundances of the two sequences shown in Fig.~2B are not equal, but occur with a highly stable ratio. This could reflect a stable difference in abundance of the bacteria they represent, but is more likely caused by differential amplification efficiency of these sequences by the PCA primers (Turnbaugh et al.,~2010; Klindworth et al.,~2013) and/or a different number of genomic 16S copies per cell (Tourova,~2003). Panels A-B show that sequence similarity need not imply ecological similarity. Finally, Fig.~2C illustrates that the converse is also true: sequences exhibiting identical time-dependence may have as little as 81\% sequence identity.

To quantify the generality of these examples, it is useful to define a measure of the ecological similarity of the bacterial subpopulations represented by two sequences. A natural candidate metric is the Pearson correlation of the measured abundance traces. Note, however, that the maximum correlation one can expect between the time traces of two sequences depends on their abundance: for low-abundance sequences Poisson sampling noise becomes non-negligible and sets an upper bound on the correlation coefficient. We therefore define the ``dynamical similarity'' of two traces as the Pearson correlation of their abundance, normalized by their maximum possible correlation $c_\mathrm{max}$, computed as the correlation of the higher-abundance time trace with a Poisson-downsampled version of itself (see SI). For sequence distance, we use the Hamming distance between sequences after pairwise alignment (see SI). With these definitions, we can present a 2D histogram of dynamical similarity vs. distance in sequence space for all sequence pairs constructed from the top 200 real sequences (Fig.~2D). As expected, most sequence pairs exhibit no significant dynamical similarity and are also far apart in sequence space, but a subset of closely similar sequences appears to display some degree of anticorrelation between the two measures. Zooming in on this region (Fig.~2E) makes this anticorrelation more apparent; however, even when restricted to the subset shown in Fig.~2E, the correlation coefficient remains weak ($R = -0.3$). Sequences separated by up to 6-7 nt (95\% sequence similarity) tend to be dynamically similar, the effect increasing for smaller distances, but this general trend is very loose and is not a reliable predictor of similarity for any particular pair. This result was not unexpected, and is frequently used in arguments against over-reliance on the 16S gene sequence (see, for example, Prosser et al.,~2007), in favor of methods providing functional information, such as shotgun metagenomics. The novelty of Fig.~2E lies in the fact that it was obtained entirely within the framework of 16S tag sequencing methodology.

\subsection{Cluster-free filtering can resolve distinct subpopulations with high dynamical similarity.}
\begin{figure*}[t!]
\includegraphics[width = 0.95 \textwidth]{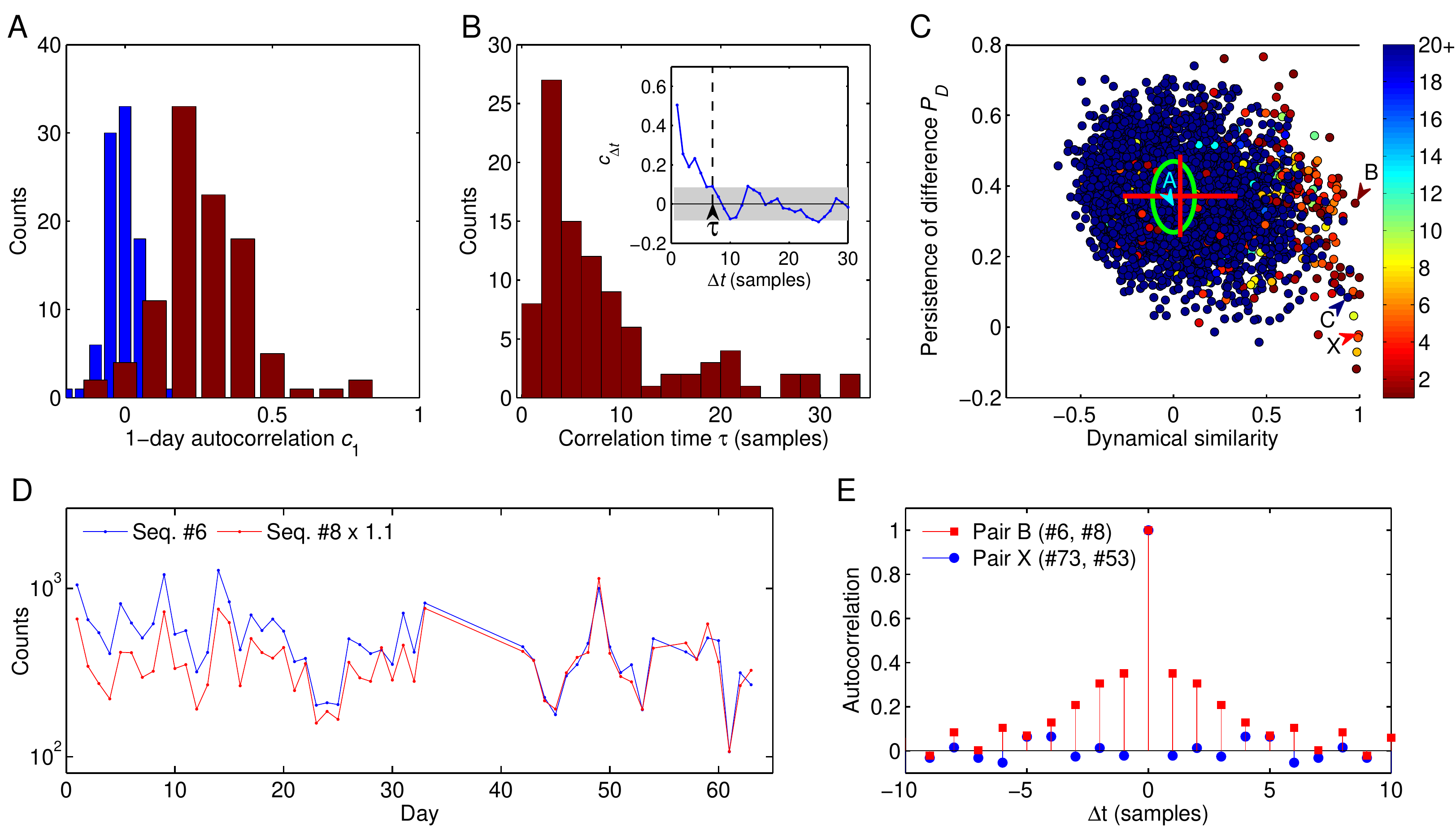}
\caption{\textbf{Dynamical similarity versus 16S similarity.}
\textbf{A.} 100 most abundant sequences of the population exhibit significant autocorrelation. Histogram of autocorrelation coefficients of sequence abundance for consecutive samples (red), and after randomly permuting sample labels (blue). \textbf{B.} Histogram of autocorrelation times of 100 most abundant sequences. We define the autocorrelation time $\tau$ as the time shift $\Delta t$ at which the autocorrelation function $c_{\Delta t}$ falls below the threshold of statistical significance as illustrated in the inset (see SI). For 19 sequences the autocorrelation time exceeds 35 days (not shown). \textbf{C.} Persistence of difference $P_D$ for all pairs of sequences from the top 100, plotted against the correlation of their abundances (normalized by maximum expected correlation $c_{\mathrm{max}}$. Green ellipse indicates mean and standard deviation for the null model obtained by reversing in all pairs the time order for one of the sequences. Most pairs are consistent with the null model, except for a broadening of the correlation coefficient distribution (mean and standard deviations indicated by the red cross). Pairs to the right of the plot are dynamically similar (strong abundance correlation), often accompanied by high sequence similarity (color code indicates Hamming distance between aligned sequences in the pair; see SI). Of these, a subset (bottom right) also exhibit weak or negligible persistence of difference. These pairs, such as pair ``X'', most likely correspond to genomic 16S variants found within a single bacterium. Letters A-C identify pairs shown in Fig.~2A-C. The large persistence of difference identifies pair B as coming from distinct bacterial cells. \textbf{D.} Sequence counts versus observation day for early samples of Seq.~\#6 and \#8 (99.2\% similarity), normalized as in Fig.~3B but excluded there due to relatively poor sequencing depth. The clear separation observed prior to day 40 confirms that these two sequences are contributed at least in part by distinct bacterial subpopulations. \textbf{E.} Autocorrelation functions of the relative difference $\Delta(t)$ for two pairs identified in (C): pair ``B'' (red squares; high $P_D$ indicative of distinct bacterial cells) and pair ``X'' (blue circles; low $P_D$ indicative of 16S variants found within a single bacterium).}
\end{figure*}

As explained in the previous section, 16S tags with low dynamical similarity clearly derive from distinct bacterial subpopulations, even if the sequences are themselves highly similar. We now consider pairs of sequences with highly correlated time traces such as observed in Fig.~2B, C. Such correlated pairs could derive from the same bacterial cells (as multiple genomic copies of the 16S gene, or as exceptionally common PCR errors not included in our model). Alternatively, they could derive from distinct bacterial subpopulations that either occupy the same ecological niche or engage in a strong obligate symbiosis. Such pairs are thus of significant ecological interest, provided it can be shown that the sequences actually derive from different bacterial cells. In this section, we demonstrate that cross-sample correlation analysis can, in some cases, successfully make this subtle distinction between same-cell or different-cell sources.

To draw this distinction, we make use of the following observation. The abundance ratio of two sequences that derive from the same bacterium is set by some sample-independent parameter (e.g.\ involving differential amplification efficiency, 16S copy number, and/or PCR error rate); therefore, any fluctuation in their abundance ratio is due to measurement noise, and must be uncorrelated between samples. Any statistically significant time (or location; see SI) correlation of abundance ratio fluctuations, e.g.\ in consecutive (or proximate) samples, is therefore strong evidence that the two sequences are at least partially contributed by physically distinct subpopulations.

For this approach to succeed, the dynamics of individual subpopulations must be slow enough to allow correlations between consecutive samples to be observed. We therefore began by computing, for each of the top 100 sequences, the autocorrelation function $c_{\Delta t}$, defined as the correlation between abundance fluctuations in samples separated by $\Delta t$ time points, and normalized so that $c_0=1$ (for simplicity, we treat samples as though they were equally spaced in time, which is approximately correct; the mean separation between samples was 1.1 days). The environment experienced by tongue microorganisms changes frequently, and one might have expected that daily sampling would probe the space of possible community states, but provide little information about community dynamics as these would occur on a faster time scale. Surprisingly, we found the time dependence of most sequences in the top 100 to have a significant autocorrelation despite the relatively low sampling rate (Fig.~3A). Although conditions on the tongue make fast abundance changes possible, as evidenced by the large, rapid fluctuations in Fig.~2A-C, we found the correlation time for the top 100 sequences to be surprisingly long, typically 2-4 days but often longer (Fig.~3B), sometimes exceeding a month (Fig.~S10).

These multi-day autocorrelations make it plausible that for physically distinct subpopulations, the fluctuations of their abundances relative to each other could be slow enough to be detectable even if their ecology is similar. Consider two sequences $A$ and $B$ whose abundance time traces are highly correlated. Denote by $n_A (t)$, $n_B (t)$ the two traces renormalized to the same mean for best overlap, as in Fig.~2B,C, and let $\Delta(t)$ be their fractional difference in a given sample (a quantity more robust to noise than the na{\"\i}ve abundance ratio):
$$
\Delta(t)=\frac{n_A-n_B}{(n_A+n_B)/2}.
$$

If $n_{A,B} (t)$ reflects abundances of two distinct subpopulations, then $\Delta(t)$ can be expected to exhibit an autocorrelation on par with that observed for the individual sequences. Intuitively, if on day 1, subpopulation $A$ is, say, 10\% more abundant than $B$, and the dynamics of both are slow, then $A$ is likely to maintain its lead on day 2. In contrast, if the two sequences are genomic variants contained within the same bacterium, then any difference between $n_A (t)$ and $n_B (t)$ must be due to measurement noise, and $\Delta(t)$ will be uncorrelated between samples. We therefore introduce the \emph{persistence of difference} $P_D$ as the 1-day autocorrelation coefficient of $\Delta(t)$:
$$
P_D=\frac{\left\langle \Delta(t) \Delta(t+1)\right\rangle}{\left\langle \Delta(t)^2\right\rangle},
$$
where angular brackets denote averaging over time. $P_D$ characterizes the persistence of abundance fluctuations of two sequences relative to each other. For sequences arising from the same cells, $P_D$ must vanish. Any pair of sequences exhibiting a statistically significant $P_D$ must be contributed, at least in part, by two physically distinct bacterial subpopulations. Note that the absolute abundance of a sequence may change dramatically between days (e.g.\ more favorable conditions can cause both subpopulations to proliferate quickly), but the normalization of $\Delta(t)$ makes $P_D$ insensitive to such overall correlated behavior.

Summarizing the above, we have the following expectation for $P_D$: For a randomly chosen pair of sequences, with insignificant dynamical similarity, $P_D$ should be significantly non-zero (due to the slow dynamics of the individual subpopulations; see SI), and form a unimodal distribution consistent with the null model of unrelated subpopulations. In contrast, pairs displaying high dynamical similarity come in two types, and the persistence of difference $P_D$ should display a bimodal distribution: pairs of sequences found within the same bacterial cell will have vanishing or insignificant $P_D$, while pairs belonging to distinct subpopulations will likely exhibit a persistence of difference comparable with the null model prediction.

This is precisely what we observe. Fig.~3C shows, for all sequence pairs constructed from the top 100 sequences, a scatter plot of their persistence of difference $P_D$ versus dynamical similarity as defined previously (the normalized Pearson correlation of their abundances). The mean and standard deviations of the distribution predicted by the null model (unrelated subpopulations) are indicated by the green ellipse, and were computed directly from the data by reversing in all pairs the time order for one of the sequences. The mean and standard deviations of the actual data are indicated by the red cross. We find, as expected, that the $P_D$ score of dynamically dissimilar sequence pairs is unimodal and consistent with the null-model prediction. In contrast, the $P_D$ score of dynamically similar pairs exhibits the predicted bimodality (right side of the plot), with a subset exhibiting weak or negligible persistence of difference (bottom right). As explained above, we interpret these low-$P_D$ pairs as corresponding to genomic 16S variants found within a single bacterium. Letters A-C identify pairs shown on Fig.~2A-C. Note that the strong persistence of difference identifies the pair ``B'' as being contributed, at least in part, by distinct bacterial cells, despite 99.2\% sequence similarity and an almost perfect correlation of abundances (Fig.~2B). Conversely, the low-$P_D$ pair ``C'' (with only 81\% sequence similarity) likely corresponds to an example of two dissimilar 16S genes contained within a single bacterium. Note the enrichment of pairs with high sequence similarity among the dynamically similar pairs, as indicated by the color code (compare with Fig.~2D).

Remarkably, in the case of pair ``B'', the conclusion of distinct bacterial subpopulations drawn from Fig.~3C can be confirmed directly. Panel D shows the time traces of this pair for days 1-64 (normalization as in Fig.~2B). Due to the relatively poor sequencing depth in these early samples, they were not included in Fig.~2B. The clear separation observed prior to day 40 provides an independent confirmation of our conclusion. We stress that these data were not used in the analysis presented in Fig.~3C, but the sensitivity of the autocorrelation method was sufficient to identify these sequences as deriving from physically distinct cells based solely on the data shown in Fig.~2B. The autocorrelation function of the fractional difference $\Delta(t)$ for this pair is shown in Fig.~3E. We have verified that the persistence of difference for this pair does not change significantly if any window of 100 consecutive samples is used instead of the full time series (data not shown).

\subsection{Clustering reads into OTUs vastly underestimates ecological richness}
Figs.~2A, S7, S10, and S11 provide examples of some fine features that standard OTU-based methods would fail to detect, but which become accessible with cluster-free filtering. We now ask whether such cases are the exception or the rule. For a given sequence similarity threshold, we can define, for each of the most abundant sequences, its would-be OTU, namely the ensemble $\{S_i\}$ of all ``real'' sequences within the chosen similarity threshold. We construct the time trace of the abundance of this OTU as the sum of the abundances of all its members. We can now ask: how representative is this time trace of the true behavior of the member sequences? Let $\{c_i\}$ be the correlation coefficients between time traces of individual members and the OTU itself, normalized to the maximum expected correlation as before. We define unweighted and weighted OTU quality scores $Q_u$ and $Q_w$ as, respectively, the simple average of $\{c_i\}$, and an average weighted by the abundance of the member:
$$
Q_u=\frac 1K \sum_i c_i      \qquad\text{and}\qquad     Q_w=\frac{\sum_i N_i c_i}{\sum_i N_i}
$$

Here $K$ is the number of subpopulations in the OTU and $N_i$ is the average abundance of member $i$. The weighted quality score $Q_w$ is always larger, because the most abundant sequence dominates the sum and so is better correlated with the OTU trace. Thus $Q_w$ tells us how representative the OTU is of its most abundant member. The unweighted quality score $Q_u$ tells us how diverse is the group of subpopulations lumped together into an OTU. If the sequences grouped into an OTU are all dynamically identical (are Poisson-resampled versions of each other at different abundances), both quality scores will be close to 1. If the OTU is dominated by one subpopulation, with other members dynamically different but very low in abundance, we will have $Q_w\approx 1$, but $Q_u\ll1$. Finally, if the OTU contains several dynamically distinct subpopulations at comparable abundances, both quality scores will be low.

The average quality scores for OTUs assembled around the top 5 sequences are presented in Fig.~4 as a function of sequence similarity threshold. The relatively high weighted quality score $Q_w$ means that an OTU time trace is, on average, fairly representative of its most abundant member. The unweighted score $Q_u$ is, however, dramatically lower, indicating that the OTUs group together sequences from subpopulations with high dynamical diversity.

\begin{figure}[t]
\includegraphics[width = 0.8 \linewidth]{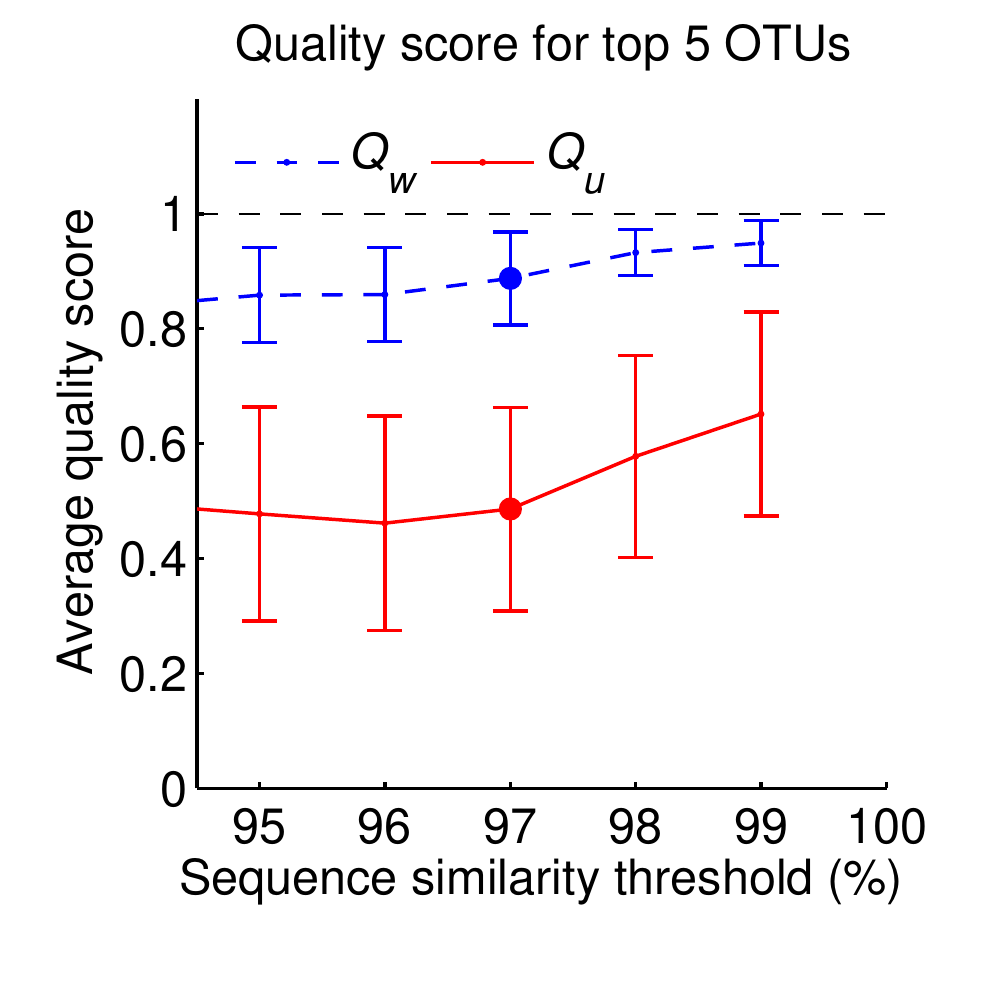}
\caption{\textbf{Clustering reads into OTUs vastly underestimates dynamical diversity.}
Average quality score for OTUs assembled around the top 5 sequences (defined as the ensemble of ``real'' sequences within a given sequence similarity threshold), as a function of similarity threshold. Error bars are standard deviations across 5 considered OTUs. Weighted quality score $Q_w$ (dashed line; see text) is high, indicating that the OTU time traces are representative of the time traces of their most abundant members. However, the unweighted score $Q_u$ (solid line) is dramatically lower, indicating that OTUs group together sequences with very different time traces. Thus OTUs combine strains with high dynamical diversity. The commonly used ``species-level'' similarity threshold of 97\% is highlighted.}
\end{figure}

These quality scores rely on abundance time-trace correlations, which become contaminated with noise for low-abundance sequences. For the purposes of Fig.~4, to apply these definitions conservatively, we therefore restricted our attention only to high-abundance members of the OTU, considering only sequences from the top 200 by overall abundance. Further, our cluster-free filtering method also has finite resolution, as the sequences we analyze are only 130nt long and may derive from distinct 16S genes, implying some unresolved diversity. This limited resolution leads to an artificial inflation of OTU quality scores as the similarity threshold approaches 100\%. For both these reasons the true quality scores of OTUs are likely even lower (see SI).

\subsection{Exact tag sequence identity is substantially more predictive of subpopulation dynamics than 99.2\% sequence similarity}

\begin{figure*}[t!]
\includegraphics[width = 0.95 \textwidth]{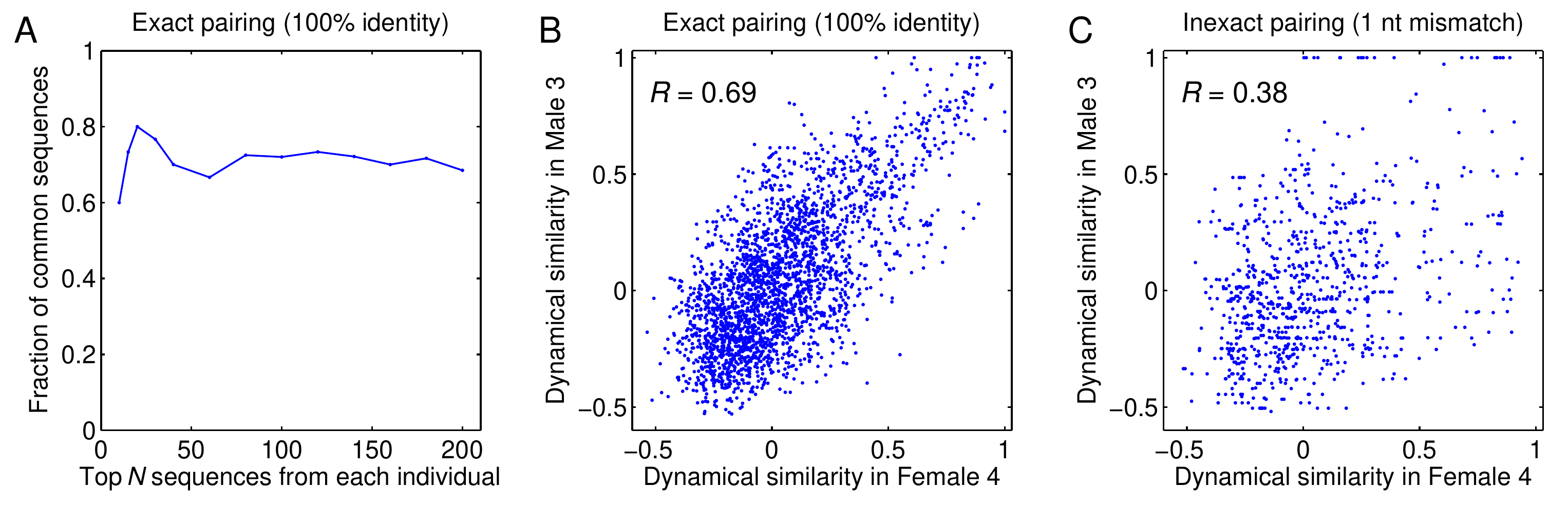}
\caption{\textbf{Comparative analysis at 100\% sequence identity of oral community composition in two cohabiting individuals reveals shared subpopulations.}
\textbf{A.} Fraction of shared 16S sequences, defined as the fraction of common tags (at 100\% sequence identity) among the most abundant $N$ sequences in each of the two individuals, plotted as a function of $N$. \textbf{B.} Scatter plot of the dynamical similarity of pairs of common sequences, as measured independently in the two individuals, for all possible pairs among the 73 common sequences shared within the top $N = 100$. \textbf{C.} Same as (B), but with intentionally inexact pairing of sequences across individuals (each sequence is mapped to a partner differing by exactly 1 nt). Despite 99.2\% sequence similarity of such pairs, allowing the 1nt mismatch significantly decreases the degree to which dynamical similarity as observed in the two individuals is correlated.}
\end{figure*}

The fact that tag sequence similarity within the 16S gene is only loosely correlated with dynamical similarity (Fig.~2E) was not unexpected (see, for example, Prosser et al.,~2007 and references therein). At a neutral mutation rate of order 10-9 per base pair per generation (Ochman,~2003), an average difference of a single nucleotide out of 100 would already require divergence for millions of generations. A more precise estimate of divergence time should take into account the possibility of horizontal gene transfer, whose rate in an ecologically relevant setting is hard to assess. However, it is clear that, generically, two bacteria that differ by even 1 nt in a particular hypervariable region of the 16S gene likely diverged a long time ago. These bacteria are likely to also differ elsewhere in their 16S gene, and to carry even more significant differences in functional parts of their genome.

In contrast, what if we consider two bacteria whose sequenced portions of their 16S genes are identical? Since the length of the sequenced fragment is small (typically $\sim100$ nt) and the mutation rate is low, these bacteria could still have diverged a very long time ago (Lukjancenko et al.,~2010). However, depending on circumstances, the actual time since the last common ancestor may be much shorter. For example, consider two communities that frequently exchange members. If two bacteria drawn from two such communities are 100\% identical in their 16S tags, a likely explanation for this identity is a recent exchange event, in which case the entire genomes of these bacteria may be close to identical. We conclude that in the presence of strain exchange between communities, exact sequence identity and near-identity may have fundamentally different implications. The study of Caporaso et al. sampled the tongue microbiota of two cohabiting individuals (Rob Knight, personal communication), and so strain exchange is likely to be a highly significant factor (Song et al.,~2013). We hypothesized, therefore, that these communities would share some non-negligible number of subpopulations at 100\% sequence identity, and that these common subpopulations might have similar ecology in both communities.

We began by identifying the fraction of common 16S sequences in the list of the top N for each individual (at 100\% sequence identity). Based on our strain exchange hypothesis, we expected to find some matches, but were still surprised to find this fraction to be as high as 75\% (Fig.~5A). Such a high proportion of perfect matches provides strong evidence that the identical sequences found in these two communities most likely diverged from a common ancestor more recently than any pair of close, but non-identical sequences within the same community. The same conclusion is supported by the analysis of fecal samples from the two individuals (Fig.~S11).

We then considered the 73 sequences that were found among the top 100 of both individuals and asked whether the behavior of these subpopulations was predominantly shaped by their presumed common origin (causing them to be similar) or by local adaptation (causing them to diverge while leaving the 16S region intact; see Lukjancenko et al.,~2010). To this end, for each pair of sequences $(i,j)$ drawn from this list, we measured their dynamical similarity independently in the two datasets;   $S_{ij}^M$ for the male and $S_{ij}^F$ for the female. If the effect of local adaptation were dominant, then the exactness of a match of 16S sequences would not carry much information: the ecologies and genomes would be no more similar between 100\%-identical partners in the two communities than between any other sequences within the same bacterial ``species'' (OTU); this scenario is implicitly assumed by taxonomy-based methods. Alternatively, if the ecology were determined primarily by the shared recent ancestor, then identical 16S tag sequences in the two communities would correspond to bacterial subpopulations with almost identical genomes. In this scenario, provided local adaptation did not modify the ecology of a subpopulation significantly, $S_{ij}^M$ and $S_{ij}^F$ should be strongly correlated, and unlike the first scenario, this correlation would be noticeably degraded for any less than 100\% sequence identity. The latter is indeed what we observe (Fig.~5B-C). Fig.~5B demonstrates that subpopulations identified by the exact same 16S tags in the two individuals are dynamically similar; see also Fig.~S11D and S12. To obtain Fig.~5C, we constructed an ``inexact pairing'' of sequences between individuals, whereupon each sequence from the top 100 in the female individual was matched to the highest-abundance sequence from the top 100 in the male individual that differed from it by exactly 1 nucleotide, when such a match existed. This matching corresponds to 99.2\% sequence identity, yet already substantially degrades the correlation between $S_{ij}^M$ and $S_{ij}^F$ (Fig.~5C). We conclude that 100\% identity of tag sequences has qualitatively different implications from even 99.2\% near-identity.

\section{Discussion}
In this work, we have demonstrated that cross-sample correlation analysis of denoised 16S data can be exploited to achieve sub-OTU resolution. The cluster-free filtering approach we presented reliably identified up to 20 distinct subpopulations within standard 97\% similarity OTUs, and a comparative analysis of oral communities of two cohabiting individuals demonstrates that most such subpopulations are shared between the two communities. Furthermore, subpopulations identified by the exact same 16S tags in the two individuals are dynamically similar, whereas even a single nucleotide mismatch is enough to degrade this similarity. Overall, our analysis shows that coarse-graining sequence data into OTUs is not essential for ecological applications of 16S tag sequencing methodology.

Our approach combines two novelties. First and foremost, we do not cluster similar sequences together. Regrettably, in the literature the term ``clustering'' has multiple meanings. Most denoising algorithms aim to assign erroneous reads to their most likely source, to make the abundance estimates of true sequences more accurate. The same term ``clustering'' is used both for this read remapping and for merging multiple true sequences into a single OTU. However, these two practices are fundamentally different. Read remapping constitutes data denoising; as such, it is always advantageous, can be done in a principled way, and can be evaluated against an objective standard of performance. Adding it to our approach would likely somewhat improve the results. In contrast, OTU clustering is a form of data coarse-graining, and the optimal degree of coarse-graining is necessarily application-dependent. Importantly, for some applications it may not be necessary or desirable. When studying coarse features of community composition and dynamics, e.g.\ comparing communities across habitats (Costello et al.,~2009, Huttenhower et al.~2012) coarse-graining is appropriate. For example, metrics of community comparison such as UniFrac (Lozupone \& Knight,~2005) are widely used precisely because, by construction, they are not sensitive to OTU sub-structure. However, when studying subtle differences between broadly similar communities, e.g.\ samples from similar habitats or repeated sampling of the same habitat, the sub-OTU structure becomes a valuable source of insight. This is the intended application for our approach. Although we focused on longitudinal Illumina data, the denoising algorithm we developed does not assume short read length or low error rate and is directly applicable to a wide range of dataset types (see examples in Fig.~S7 and S11), provided the error structure is consistent across samples (Preheim et al.,~2013). We expect our approach to be useful for investigating the structure and dynamics of discrete community subtypes such as those observed in the vaginal community (Huttenhower et al.~2012).

Our second novelty is to exploit the quantitative advantage offered by multi-sample (time-course or cross-sectional) data. Since the copy number of the 16S gene carried by a bacterium is typically unknown (Tourova,~2003), and the PCR amplification bias among different 16S fragments can sometimes reach orders of magnitude (Turnbaugh et al.,~2010; Klindworth et al.,~2013), the 16S data from a single sample carries very little quantitative information about community composition. In contrast, the ratios of sequence abundance  are highly informative and can be measured very precisely, as demonstrated in Fig.~2B,C. Recently, time-course data collection has been gaining in popularity, as it was recognized that such experiments can offer valuable insight into community dynamics (Shade et al.,~2013 and references therein). However, another major advantage of such datasets, namely that changes in sequence abundance ratios can be measured much more accurately than absolute abundances, is only beginning to be explored. For us, time-series data provides a context where sub-OTU resolution acquires its full power. Specifically, we have shown that cross-sample comparisons enable us to decouple sequence similarity from dynamical similarity while remaining fully within the framework of 16S tag sequencing. High-quality reference databases can complement our approach to facilitate paralog identification. The basic methodology described here should also be extendable to other marker genes.

The new approach described in this work is not a replacement for OTU clustering; it discards low-abundance sequences and so is unsuitable for studies of population-level alpha- or beta-diversity. However, the novel statistical and computational techniques we present allow full utilization of the quantitative information carried by sequences with a moderate-to-high abundance. This has promising applications for the study of factors affecting community assembly. As discussed above, sub-OTU resolution can provide insight into the prevalence of strain exchange between communities, invasion / extinction dynamics of OTU subpopulations, and the time scale of ecological divergence relative to sequence divergence. In addition, the dynamics of individual-specific subpopulations could help characterize the role of host genetics or the host immune system on shaping the community, particularly in the context of highly controlled experiments with germ-free animals.

\section*{Acknowledgements}
We thank W Bialek, AF Bitbol, CP Broedersz, DS Fisher, R Knight and members of his lab, SA Levin, Y Meir, SW Pacala, SP Preheim and MJ Rosen for helpful discussions, G Caporaso for providing the data, and R Edgar for USEARCH support. This work was partially supported under the DARPA Biochronicity program, Grant D12AP00025, and National Science Foundation Grants PHY-1305525 and CCF-0939370.

\def\bibindent{1em}

\endgroup

\appendix
\cleardoublepage
\onecolumngrid

\section*{Supplementary Information}
 \renewcommand{\tabcolsep}{0.5cm}
 \setcounter{figure}{0}
 \renewcommand{\thefigure}{S\arabic{figure}}
 \renewcommand{\thetable}{S\arabic{table}}
\tableofcontents

\section{Supplementary methods. Cluster-free filtering: details and applications}
In this section, we illustrate the idea of error-model-based denoising (see also the introduction in Rosen et al., 2012) and give a detailed description of the simple denoiser we designed for this work. We then describe the workflow of an open source software package we created to implement this denoiser, and compare its performance on mock community data with DADA (Rosen et al., 2012). Finally, to illustrate that our cluster-free filtering approach is not restricted to longitudinal Illumina data, we provide an example of its application to a very different dataset, specifically 454 sequencing data from a cross-sectional environmental sampling performed by Preheim et al., 2013.

\subsection{Motivation: sequencing noise is low}
Clustering can be a useful strategy for filtering noise by coarse-graining data. However, such coarse-graining may not be a necessity: if the noise level is low, as suggested by known estimates of PCR and sequencing error rates (see, for example, Quince et al. 2011), then we can avoid clustering, since we expect each community member to be predominantly represented by the same 16S sequences.

We begin by illustrating this idea using the tongue microbiome data of Caporaso et al. Since the tongue community is relatively stable (Costello et al., 2009), the low-noise scenario would predict that certain specific sequences should consistently dominate in each sample. Alternatively, if the noise were high, then the high-abundance community members would be represented by clouds of similar reads, none of which would clearly dominate.

\begin{figure}[h!]
\includegraphics[width = 0.4 \textwidth]{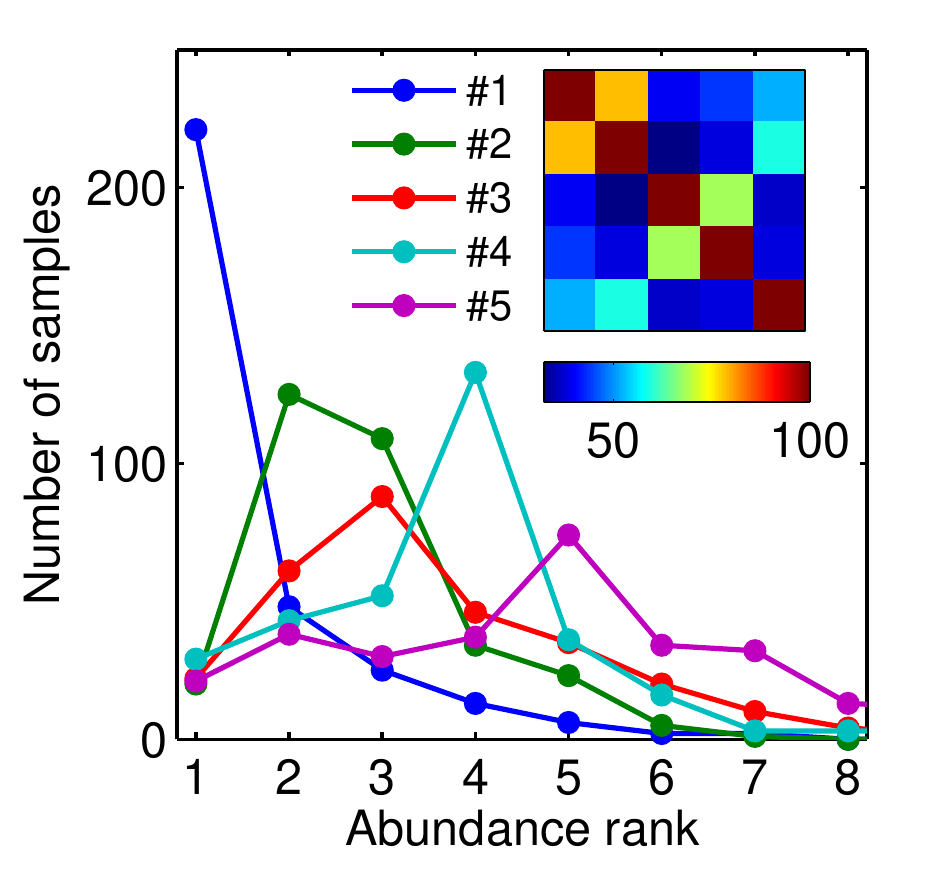}
\caption{The distribution of ranks for the top 5 sequences over all samples. Inset: pairwise sequence similarity (\%). The top sequences are strongly distinct and their rank is consistent across samples.}
\label{fig:Rank}
\end{figure}

To show that the data of Caporaso et al. supports the first (low-noise) scenario, we identified the top 5 sequences by overall abundance. These sequences were strongly different (Fig.~\ref{fig:Rank}, inset), corresponding for the most part to bacteria from different phyla: in decreasing order of abundance, these were \textit{Neisseria sp.} (phylum \textit{Proteobacteria}, class \textit{Betaproteobacteria}), \textit{Haemophilus sp.} (phylum \textit{Proteobacteria}, class \textit{Gammaproteobacteria}), \textit{Fusobacterium sp.} (phylum \textit{Fusobacteria}), \textit{Streptococcus sp.} (phylum \textit{Firmicutes}), and \textit{Prevotella sp.} (phylum \textit{Bacteroidetes}). (Taxonomy assigned by a BLAST search (BLASTN 2.2.22, matrix $(1, -1)$, gap/extenstion penalty $(5, 2)$) against GreenGenes database; DeSantis et al., 2006. All 5 sequences had a match with 100\% identity over 100\% of sequence length.) The sample-by-sample rank of these overall top 5 sequences was consistently in the top 10. We stress that the goal of Fig.~\ref{fig:Rank} is not to characterize the temporal stability of community composition (previously characterized, for example, in Costello et al., 2009); rather, it serves to show that the community members that correspond to these highest-abundance tags are consistently represented by the same 130nt sequence (at 100\% identity) across all samples. In other words, despite the presence of noise in the data, 100\% sequence identity is not an unreasonable criterion: the error rate is low enough that the error-free sequence dominates over the ``error cloud'' of its variants (Edgar, 2013). This key observation is the foundation of the approach described in this work.

\subsection{Estimating rates of one-nucleotide substitutions}
\begin{figure}[b!]
\includegraphics[width = 0.95 \textwidth]{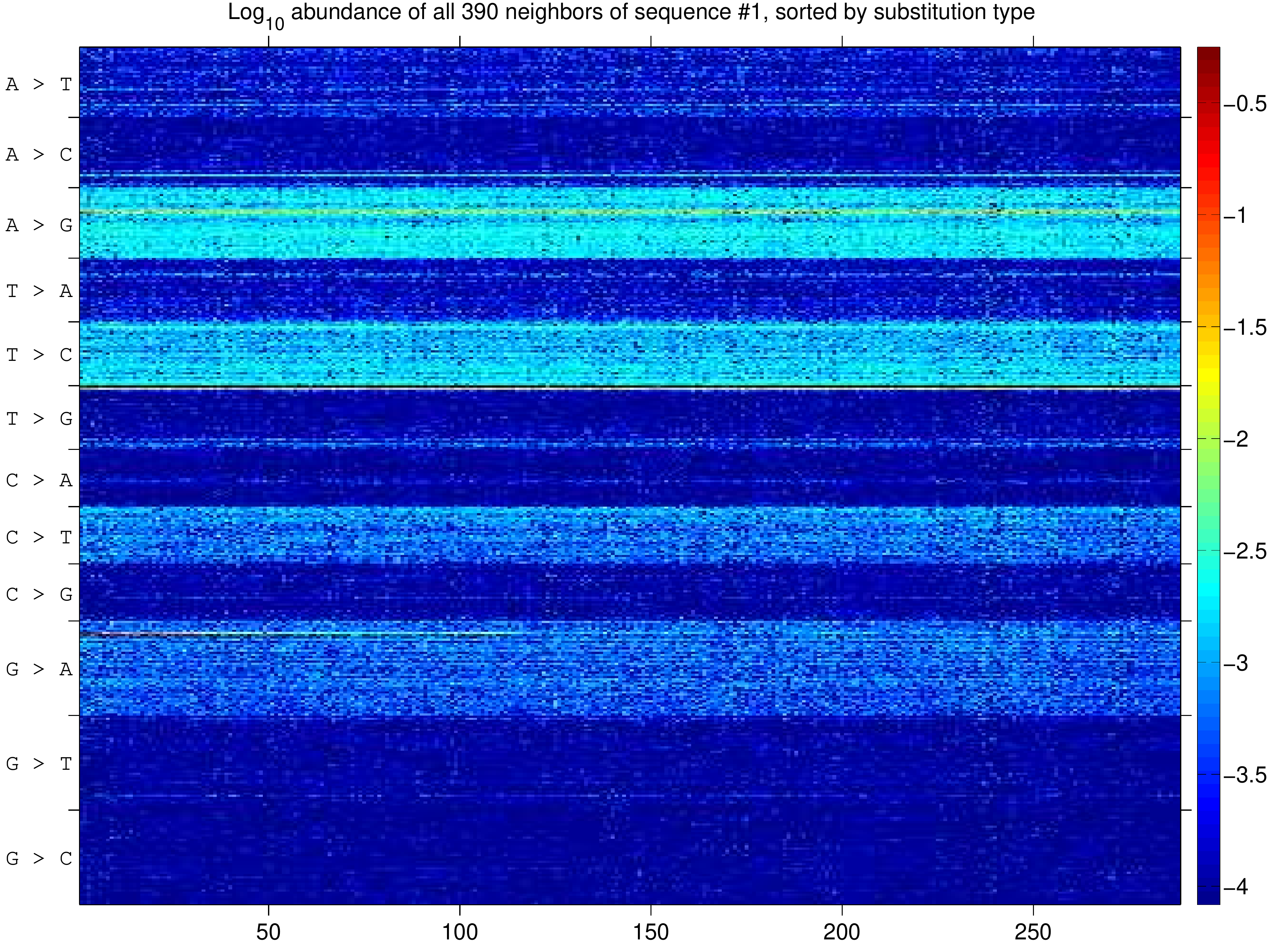}
\caption{This extended version of Fig.~1B shows all 390 first neighbors of Seq.~\#1, ordered by the type of substitution (and within these classes, by the position of the substitution along the sequence). Color indicates abundance on a log scale, normalized to the abundance of Seq.~\#1. Except for a few overrepresented neighbors (\emph{cf.} Fig.~1B), the substitution type accounts for most of the variance in neighbor abundance.}
\label{fig:ErrorCloud}
\end{figure}

To estimate the rates of substitution errors observed in data after quality filtering, we used the ``error clouds'' around the high-abundance sequences in the dataset. Since all sequences were trimmed to a length of 130nt, each ``mother'' sequence has 390 direct neighbors in sequence space ($\text{Hamming distance} = 1$). For very high-abundance sequences such as Seq.~\#1, all 390 neighbors were observed in at least one sample of the time series. The time series of their abundances, normalized to the abundance of Seq.~\#1, is shown in Fig.~\ref{fig:ErrorCloud}. For this figure, the neighbors were ordered by the type of substitution that differentiates them from the mother sequence, and, within these categories, by the position of the differing nucleotide along the sequence. We see that, with a few exceptions (most notably the three neighbors also shown in Fig.~1B), the abundance of a given neighbor is a constant fraction of the abundance of the mother sequence. This is precisely what we expect for neighbors that arise as PCR or sequencing errors of the mother sequence, and the abundance ratio is then the probability of that particular error.

We see that the error rate is set primarily by the type of substitution, and does not exhibit significant dependence on the position along the sequence. For long reads, we would likely have seen an increase in error rates towards the end of the sequence, but our sequences are only 130nt long, well within the capabilities of accurate base-calling of the Illumina platform. We can therefore assign probabilities to substitution errors based solely on the substitution type (which nucleotide was replaced by which other), independent of the position along the read.

To determine these probabilities, we first identify the neighbors that are outliers in their substitution category; they likely correspond to true biological sequences physically present in the community, rather than sequencing errors. Outlier exclusion is done based on z-scores, i.e. for each sequence we compare its raw cumulative abundance over all samples to the mean in its substitution category, and normalize by the standard deviation in the category. A strong outlier differing from the mother sequence by a nucleotide substitution at location $K$ will skew the error rate estimation at that location: some substitution type will appear to be unusually frequent. Therefore, we exclude nucleotide locations that correspond to the strongest outliers, those for which the z-score exceeds some threshold. The remaining locations are then used to estimate the error rates: for each of these locations, we count the number of times a particular substitution occurred, as well as the number of times the nucleotide was recorded correctly. After appropriate normalization, these counts give us the probability of each type of the error.

\begin{figure}[t!]
\includegraphics[width = 0.5 \textwidth]{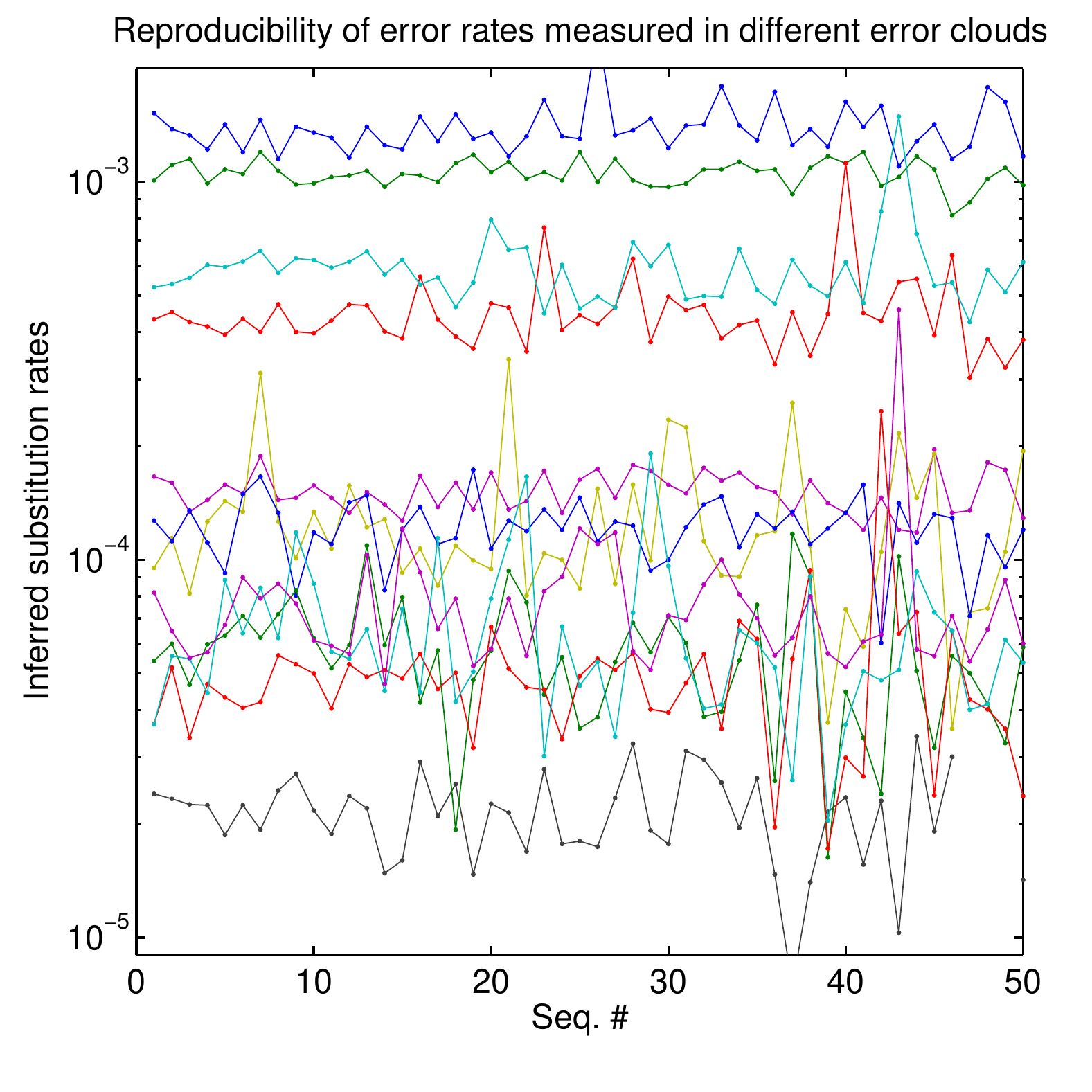}
\caption{The substitution error rates directly inferred from the error clouds of top 50 sequences by abundance are reproducible across error clouds. Each of 12 separate plots shows the inferred rate of a specific substitution (not labeled to reduce clutter; see also Fig.~\ref{fig:QCparams} and Sup. Table S1). Predictably, the variability increases when the error clouds of less abundant sequences are used.}
\label{fig:ErrorRates}
\end{figure}

Fig.~\ref{fig:ErrorRates} shows the inferred substitution rates for error clouds around the top 50 sequences by abundance, using minimally quality-filtered data processed as described in the Methods (Phred score cutoff 2, z-score threshold 2). We find that the rates of different substitutions can differ dramatically (up to ~50-fold), but our estimates are highly reproducible (note the log scale on the Y axis), with variability predictably increasing if lower-abundance error clouds are used. Table~S1 lists the error rates estimated from the error clouds of the top 10 sequences (mean $\pm$ standard deviation). Note that, in principle, this effective error probability includes both the base-call errors of the Illumina sequencer and the single-nucleotide substitution errors occurring during PCR. However, the approximate symmetry between rates of a substitution and its reverse-complement partner (e.g. $p_{T\rightarrow A}\approx p_{C\rightarrow G}$), and a clear bias towards transitions as opposed to transversions, suggests that the observed substitutions are dominated by PCR errors (compare with Quince et al., 2011, Table 2).

\begin{figure}[t!]
\includegraphics[width = 0.55 \textwidth]{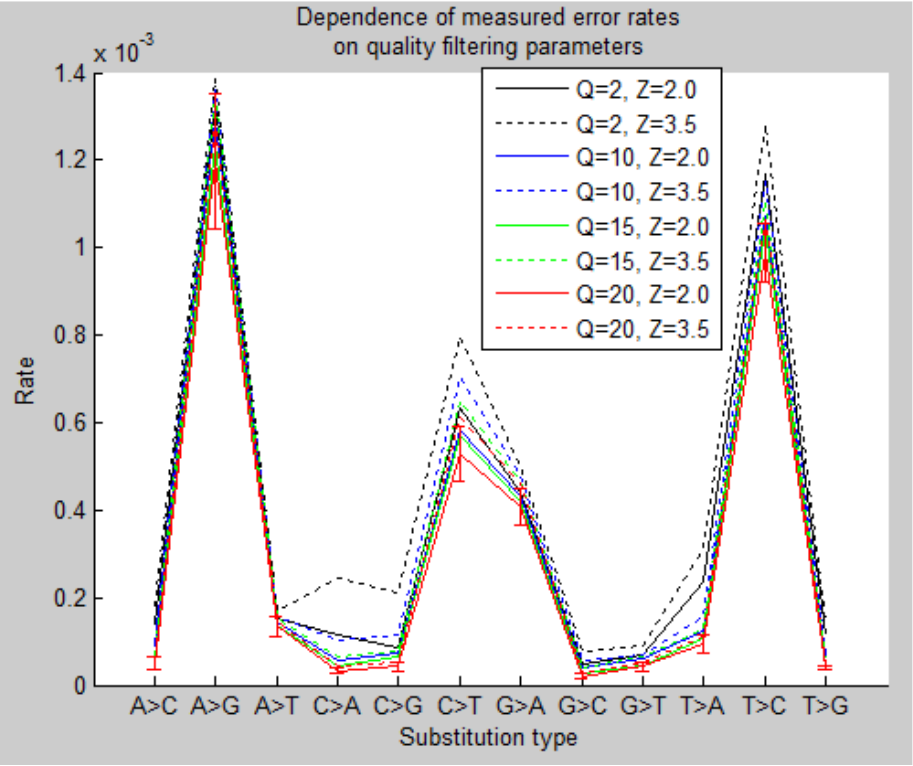}
\caption{The estimates of substitution error rates exhibit only a weak dependence on Phred quality score filtering parameters, as expected if the dominant source of substitution errors is PCR amplification rather than base call errors. Shown are average error rates as measured for top 10 sequences in the sample. Error bars on the (Q=20, Z=2) plot indicate standard deviation across the 10 estimates.}
\label{fig:QCparams}
\end{figure}

Inferring error rates directly from the data offers multiple strong advantages. Specifying the error rate as an external parameter (e.g., Morgan et al., 2013) necessarily requires resorting to a conservative global upper bound. Different PCR conditions and different sequencing machines will have different error rates (for example, compare Fig.~\ref{fig:QCparams} and Fig.~\ref{fig:DBC}A). Further, substitutions vary strongly in probability: in our case, using a single upper bound on error rates would have over-estimated the probability of certain error types by up to 50-fold, reducing our ability to resolve close sequences. In other words, measuring substitution rates directly from the data both reduces the number of algorithm parameters and improves performance.

\begin{table}[t]
\centering
\begin{tabular}{|l|c|c|c|c|c|}
\hline
                & $\rightarrow A$ & $\rightarrow C$ & $\rightarrow T$ & $\rightarrow G$ & Total \\
\hline
$A\rightarrow$  &                 & $0.14\pm0.06$   & $0.15\pm0.02$   & $1.34\pm0.12$   & $1.63\pm0.14$ \\
\hline
$C\rightarrow$  & $0.07\pm0.02$   &                 & $0.59\pm0.04$   & $0.06\pm0.01$   & $0.72\pm0.05$ \\
\hline
$T\rightarrow$  & $0.12\pm0.03$   & $1.06\pm0.07$   &                 & $0.07\pm0.01$   & $1.26\pm0.08$ \\
\hline
$G\rightarrow$  & $0.42\pm0.03$   & $0.02\pm0.01$   & $0.05\pm0.01$   &                 & $0.49\pm0.03$ \\
\hline
\end{tabular}
\caption{Substitution error rates per nucleotide, multiplied by 1000, as measured from the ``error clouds'' of the top 10 sequences by abundance. Error bars are standard deviations across the 10 estimates.}
\end{table}

To investigate how the measured substitution probabilities depend on the quality filtering parameters, we applied the same analysis to data filtered using different Phred quality score thresholds ($Q_{\mathrm{min}}=2,10, 15, 20$) as well as different z-score thresholds (2.0, 3.5). The results are presented in Fig.~\ref{fig:QCparams}. As expected, the average error rates increase as the Phred score threshold is lowered; however, the magnitude of this change is very small, comparable with the variability of error rate estimates across the top 10 sequences as indicated with the error bars on the plot corresponding to the most stringent filtering, $Q_{\mathrm{min}}=20, Z=2$. This provides further evidence that the majority of substitution errors occur during PCR amplification rather than during sequencing, and thus are not captured by Phred quality scores. We conclude that strict Phred quality filtering unnecessarily reduces data quantity while only marginally improving its quality; for our analysis, we therefore subjected the reads to minimal quality filtering as described in the Methods.

The dependence on z-score threshold is also consistent with our expectations: a high z-score threshold increases the error rate estimate. Predictably, including stronger outliers ($z=3.5$) causes the measured error rate to vary significantly across filtering conditions; we used $z=2$ which provided excellent reproducibility.

The reproducibility of error rates as observed on Fig.~\ref{fig:ErrorRates} justifies a posteriori our simplifying assumptions such as neglecting the probability of double substitutions in our calculation. Note that, according to the Table S1, the average total error rate per nucleotide is only $1.0$~$10^{-3}/\text{nt}$. Therefore, within our error model, assuming that errors occur independently, we estimate find that a 130~nt-long sequence has 88\% probability of being recorded with no errors. In practice, errors appear to correlate and the true zero-error probability is likely lower. As a different estimate, we calculate the total abundance of all sequences retained by our filtering (7~057~860 reads in 507 samples), and compare to the total number of reads before filtering (8~685~722 reads distributed across 1.4M unique sequences). We find that the filtering algorithm retained 81\% of all reads. Since the algorithm intentionally disregards true sequences with low abundance, this estimate is conservative. Further, this estimate is largely insensitive to the error independence assumption: given our stringent filtering criteria, even an unexpectedly frequent double error will be discarded, provided it is less common that a single error. We conclude that $>81\%$ of reads in the dataset had no errors, which justifies our decision to discard noisy reads rather than attempting to remap them to their most likely source. For longer reads or noisier data, our approach remains applicable without changes; however, the fraction of error-free reads will be lower. In this case, to avoid significant loss of sequencing depth, we recommend replacing our simple denoiser by an algorithm such as DADA that performs read remapping.

\subsection{The algorithm for filtering substitution errors}
\begin{figure}[t!]
\includegraphics[width = 0.5 \textwidth]{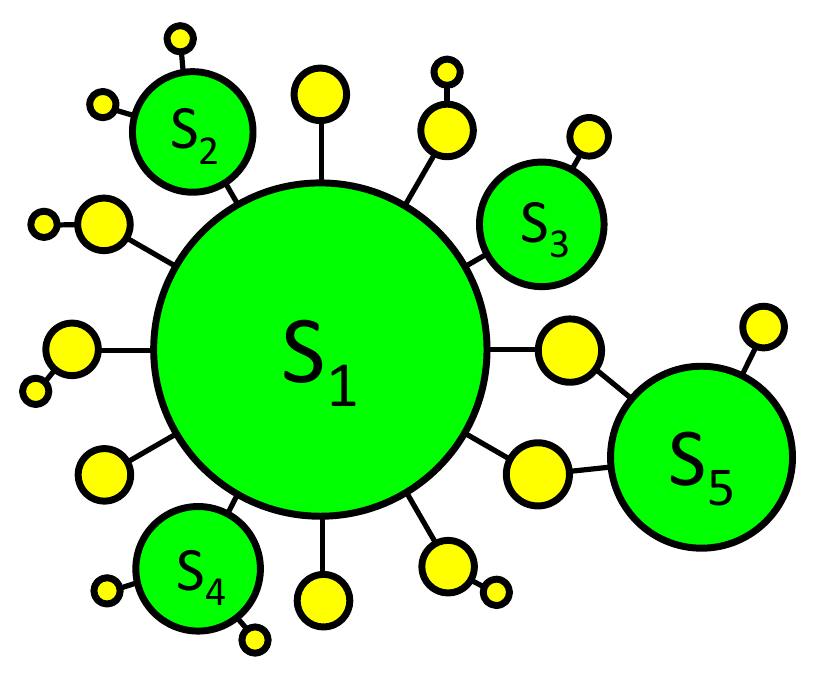}
\caption{A more detailed version of the error cloud cartoon in Fig.~1C. Each circle is a unique sequence, with size representing abundance in a sample. True biological sequences ($\SSS_1$-$\SSS_5$; green circles) generate ``daughter'' variants due to substitution errors (yellow circles). Black lines denote $\text{Hamming distance} = 1$ in sequence space. The error rates calculated from the error clouds (see Fig.~\ref{fig:ErrorRates}) can be used to calculate, for every sequence, its expected abundance under the assumption that it arose through substitution errors from its more abundant neighbors. Sequences whose abundance is significantly above this expectation are labeled as real (green circles). Note that sequences may arise as substitution errors of multiple ``mother'' sequences: common neighbors of $\SSS_1$ and $\SSS_5$ in this cartoon will have a larger abundance than other substitution errors of either $\SSS_1$ or $\SSS_5$. However, if this increase in abundance is consistent with the null model, they will be correctly recognized as substitution errors.}
\label{fig:Cartoon}
\end{figure}

For the Illumina sequencing platform, substitution errors account for the bulk of the errors. As described above, these errors have a reproducible structure and their rates can be estimated directly from the data. Using these numbers, for any sequence $\SSS$ present in a given sample, we can estimate its null model abundance, denoted $N^0$ (abundance derived from sequencing errors of its more abundant neighbors), as follows (Fig.~\ref{fig:Cartoon}):
\begin{enumerate}
\item	Order sequences by decreasing abundance: $\SSS_1$, $\SSS_2$, etc.
\item	Set $N_i^0=0$ for all $i$
\item	For each sequence $\SSS_i$ with abundance $N_i$:
\begin{enumerate}
\item	Find all $j$ such that $S_j$ is a first neighbor of $S_i$ and $N_j<N_i$.
\item	For each $j$, use the substitution error table to determine the probability $p_{ij}$ of $\SSS_i$ to be recorded as $\SSS_j$
\item	Set $N_j^0=N_j^0+p_{ij} N_i$ (``spillover'' from $\SSS_i$ into $\SSS_j$)
\end{enumerate}
\end{enumerate}

This zero-parameter algorithm assigns, for each sequence, its null-model abundance expected in that particular sample, using error rates estimated directly from the data. We next use this information to identify ``candidate sequences'': those whose presence cannot be explained by a substitution-only error model. Candidate sequences are selected through an abundance criterion, requiring their abundance to exceed the null model prediction ($N^0$) by at least ten-fold, and be no less than 10 counts. We then retain all sequences that independently passed this stringent filtering in at least 2 samples. The reasoning behind this strategy is explained in the Methods section of the main text.

\subsection{Other error types, including chimeras and PCR indels}
With Illumina sequencing, substitution errors account for most of the erroneous sequences in the data, and their occurrence appears to be adequately described by a simple quantitative model. This type of errors is therefore well-suited for error-model based denoising. The list of sequences retained after denoising includes true biological sequences, but also errors not described by our model. The latter category includes chimeras, PCR indels, and possibly other errors such as context-dependent PCR substitutions occurring much more frequently than expected within our model.

We are not aware of any quantitative model for PCR indel errors, which, in our experience, are strongly context-specific. Following Rosen et al., one could make the conservative decision that whenever two candidate sequences differ by pure indels, the lower-abundance should be treated as a possible error. A corresponding script is included in our cluster-free filtering pipeline. However, by definition, this makes it impossible to resolve true biological sequences differing by an indel. Retaining putative indel errors and comparing their abundance distribution across samples with their presumed ``mother sequences'' would allow identifying such cases. Since PCR indels are comparatively infrequent, for Illumina sequencing we consider indel filtering an optional step of the pipeline. In contrast, the 454 sequencing platform introduces frequent indel errors at homopolymer regions of the sequence. For 454 data, therefore, proper indel treatment becomes a necessity. The indel-filtering script we provide offers one solution; however, since errors we seek to eliminate occur during PCR, while indels occur during 454 sequencing, the best indel treatment strategy for the 454 platform is to merge sequences into ``indel families'' (Rosen et al., 2012) prior to denoising. Implementing this functionality within our software package will improve its support of 454 data; at the moment, the better approach is to apply our cross-sample analysis to the output of the DADA denoiser (Rosen et al., 2012).

As for chimeras, in our analysis pipeline, we filter chimeric sequences with UCHIME de novo (Edgar, 2011). Following Robert Edgar (UCHIME documentation), we recommend applying chimera filtering to pooled data across samples.

\subsection{Cluster-free filtering software package}
\begin{figure}[h!]
\includegraphics[width = 0.65 \textwidth]{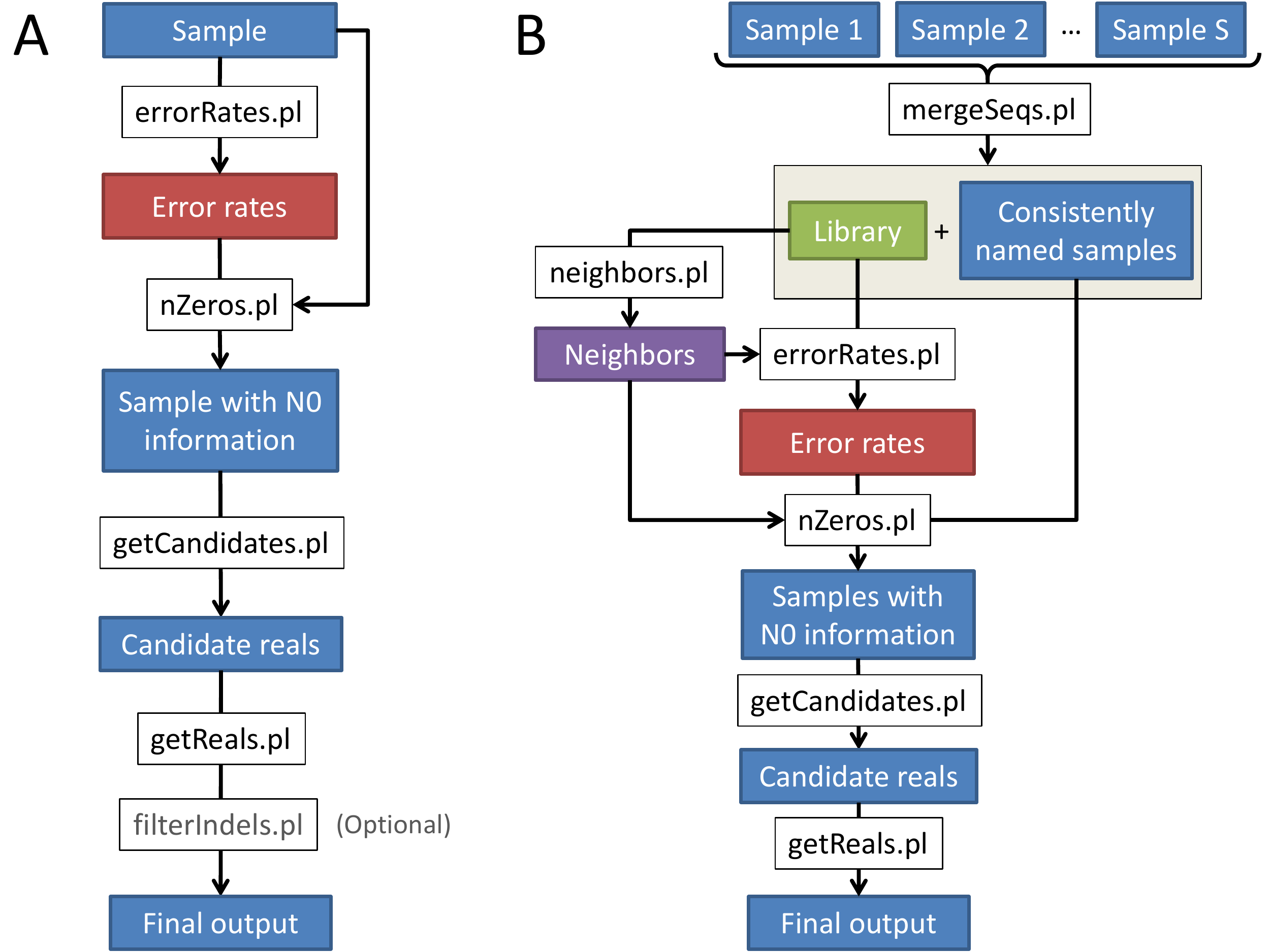}
\caption{The workflow of cluster-free filtering software package. \textbf{A:} The simplest way of running cluster-free filtering denoiser on a single sample. \textbf{B:} Extended workflow diagram appropriate for large multi-sample datasets. The optional indel filtering step is omitted for simplicity. In both cases, the blue rectangles represent dereplicated FASTA files with sequences of identical length. \textbf{getReals.pl} includes chimera filtering (performed with UCHIME).}
\label{fig:Workflow}
\end{figure}

The implementation of the denoising algorithm described here is freely available at http://github.com/hepcat72/cff as a suite of open-source Perl scripts. Fig.~\ref{fig:Workflow} summarizes the workflow of the filtering process.

Applying the denoiser on a per-sample basis is a straightforward four-step process, optionally supplemented by indel filtering (Fig.~\ref{fig:Workflow}A). However, our denoiser is specifically designed to be run on large multi-sample datasets. The extended workflow appropriate for large datasets (Fig.~\ref{fig:Workflow}B) has three key differences:
\begin{enumerate}
\item The original samples are pooled to construct a library of all unique sequences ever observed, and sequences in the original samples are renamed so that the same sequence has the same identifier in all samples (\textbf{mergeSeqs.pl}).
\item The error rates are estimated using the pooled data from all samples (i.e. the library) for better accuracy.
\item The neighbor structure is constructed once, for all sequences in the library (\textbf{neighbors.pl}). In the per-sample workflow (Fig.~\ref{fig:Workflow}A), \textbf{neighbors.pl} is automatically invoked for every sample in a manner transparent to the user, which simplifies the workflow; however, many sequences are shared across samples, so for sufficiently large datasets, explicitly calculating the neighbor structure only once results in better performance.
\end{enumerate}

The optional indel-filtering step uses MUSCLE aligner (Edgar, 2004) with modified gap penalty parameters, appropriate for detecting 454 homopolymer indels (\verb"-gapopen -400 -gapextend -399"; see documentation).

The software package is provided with built-in documentation, a test dataset and two shell scripts that allow running the entire workflow presented above with a single command: \verb"run_CFF_on_FastA.tcsh" and \verb"run_CFF_on_FastQ.tcsh" (the latter script uses USEARCH to perform the minimal quality filtering as described in the Methods). The flexible and thoroughly documented command-line interface makes it easy to incorporate cluster-free filtering into any existing pipeline. To reproduce our analysis of the data from Caporaso et al. (2011), download the quality-filtered data published with that study (available at MG-RAST:4457768.3-4459735.3), place it in a folder \verb"CaporasoData" and run:

\begin{center}
\verb|tcsh run_CFF_on_FastA.tcsh 130 analysisResults "CaporasoData/*.fna"|.
\end{center}

\subsection{Mock community validation and comparison with DADA}
To validate the performance of our simplified denoiser, we compared it with a state-of-the art denoiser DADA (Rosen et al., 2012) using two mock community datasets (\textit{Divergent} and \textit{Artificial}; Quince et al., 2011) that Rosen et al. used to demonstrate DADA's superior accuracy to AmpliconNoise. Quoting from the original publication, these datasets were constructed by amplifying the V5 region of the 16S rRNA gene from 23 and 90 clones, respectively, isolated from lake water. The \textit{Divergent} clones were mixed in equal proportions and are separated from each other by a minimum nucleotide divergence of 7\%, while the \textit{Artificial} clones were mixed in abundances that span several orders of magnitude, with some of the clones differing by a single-nucleotide substitution. For purposes of comparison, we used the exact same sets of filtered reads ($35\,190$ reads in \textit{Divergent} set; $31\,867$ in \textit{Artifiical}), kindly provided to us by Michael Rosen.

The comparison of denoiser output and the reference set of Sanger clones was complicated by the imperfections of the reference set. A number of ``reference'' Sanger clones differed from their closest high-abundant matches in the 454 data at the same locations towards the beginning of the read, which is suggestive of errors in the reference sequences. Further, some reference sequences of the \textit{Artificial} set had no close matches in the data; some Sanger clones differed at locations that were not part of the 454 sequenced fragments; and 454 sequences included 6 extra bases at the beginning of the sequence that were absent from the Sanger clones.

We therefore began by constructing ``cleaned'' reference sets as follows: for each reference Sanger clone, we found its closest match in the dataset that had at least 98\% similarity and an abundance of at least 10 counts. This matching 454 read was used as the new reference sequence, and the differences, if any, were ascribed to Sanger clone errors. For the \textit{Divergent} dataset, each reference sequence had exactly one clear match in the 454 data. For the \textit{Artificial} set, of the 90 reference Sanger clones, we found that one was 29 nts away from the closest 454 read; for 3 other clones, no 454 read within $\ge98\%$ sequence similarity radius reached an abundance of 10 counts. Our algorithm intentionally disregards any sequences below this abundance threshold; therefore, for the purposes of this comparison these reference sequences were considered absent and we did not count them as false negative for any of the algorithms. Several groups of clones were not distinguishable by the 454 sequenced fragment. Altogether, the new reference set of sequences that were both present and distinct contained 49 reference sequences.

We then ran DADA and cluster-free-filtering on both datasets. DADA was run with the same parameters as used for this data in the original publication, namely $\Omega_a=10^{-40}$ and $\Omega_r=10^{-3}$. Cluster-free-filtering included indel filtering step, since this data was obtained using the 454 platform and indels appear frequently.

\begin{table}[t]
\centering
\begin{tabular}{|l|c|c|c|}
\hline
Category                & DADA   & CFF & Abundance\\
\hline
\hline
\textit{Divergent}: 23 reference sequences & 23 true positives & 23 true positives & 231-1426 counts\\
                                           & 0 false negatives & 0 false negatives &          \\
\hline
Other detections & 0  & 0  &          \\
\hline
\hline
\textit{Artificial}: 49 reference sequences & 48 true positives & 49 true positives & 18-3587 counts\\
                                            & 1 false negative & 0 false negatives &          \\
\hline
Other detections & Seq. \#35 & Seq. \#35 & 163 counts\\
                            &           & Seq. \#95 & 13  counts       \\
                            &           & Seq. \#103 & 12 counts        \\
                            &           & Seq. \#119 & 12 counts        \\
\hline
\end{tabular}
\caption{Comparison of DADA and cluster-free filtering (CFF) denoiser on mock community data. Sequences numbered by decreasing abundance in the dataset.}
\label{tbl:mock}
\end{table}

The results are presented in Table.~\ref{tbl:mock}. Both algorithms identified correctly all 23 reference sequences of the \textit{Divergent} dataset. For the \textit{Artificial} set, and due to the conservative parameters recommended by Rosen et al., one of the reference sequences was missed by DADA but was correctly identified by our algorithm. Sequence \#35 (in order of decreasing abundance), absent from the reference set, was retained by both algorithms and is likely a true biological sequence. Cluster-free filtering generated 3 additional detections just above its threshold of 10 counts. It is instructive to trace the origin of these calls. For example, Seq.~\#95 was discarded by DADA as possibly an erroneous read generated by its closest reference sequence (Seq.~\#1) two substitutions away. Specifically, Seq.~\#95 differs from Seq.~\#1 by a T at location 23 and a G at location 118, a relation that we denote ``Seq.\#95\,=\,Seq.\#1~23T~118G''. If it were true that Seq.~\#95 is a substitution error of Seq.~\#1, we would generally expect single-error variants to be more abundant than double errors. In reality, Seq.\#1~23T (=Seq.~\#587) and Seq.\#1~118G (=Seq.~\#121) have abundances of just 4 and 12 counts, respectively, which is why our algorithm identified Seq.~\#95 as likely real. However, its unexplainably high abundance could also have arisen through amplification of a double substitution that occurred early in the PCR cycle, and the default parameters of DADA were chosen conservatively so as to eliminate such cases (Rosen et al., 2012). Whether or not these detections are false positives or true biological contaminants can be determined only by a cross-sample analysis as presented in the main text.

\subsection{Runtime comparison with DADA}
The methodology presented in this work was designed to perform cross-sample comparisons of sequence abundance in individually denoised samples. As explained in the Methods, the simplified denoiser we developed is meant to maximize performance on large datasets specifically for this application, taking advantage of our focus on moderate-to-high abundance sequences. Other denoisers can be used. To estimate the runtime of DADA on the tongue dataset considered here, we used a representative subset of 20 samples, 10 from lane 5 and 10 from lane 6 of Caporaso et al. Following the instructions in Rosen et al., 2012, we used ESPRIT to precluster sequences in each sample prior to processing them with DADA. The measured runtime is presented in Table~\ref{tbl:DADAtime}. Extrapolation to the full set of 507 samples yields the estimate of $2.3\,10^5$ sec total runtime quoted in the text, compared to 626 sec actual runtime for cluster-free-filtering. As explained in the main text, one of the reasons for this speedup is that our multi-sample detection strategy allows us, in any given sample, to look for candidate sequences only among those with abundance $\ge$10 counts. This speedup can be applied to DADA as well; to this end, we removed all clusters that contained no sequences with abundance $\ge$10 counts, and measured DADA runtime after this filtering; this decreased the estimated runtime on the full dataset to $5.5\,10^4$ sec. Using DADA in this way is the strategy we recommend for applying our cross-sample comparison methodology to 454 data with long reads where erroneous read remapping and indel family merging become advisable. Eliminating low-abundance sequences leads to a considerable improvement of DADA runtime; nevertheless, the total runtime remained two orders of magnitude slower than our cluster-free filtering approach, due primarily to the computational cost of preclustering.

\begin{table}[t]
\centering
\begin{tabular}{|l|c|c|c|c|}
\hline
                    & ESPRIT+DADA  & ESPRIT+DADA, abundant clusters only & CFF denoiser\\
\hline
Lane 5, 10 samples  &  969 + 1297 sec       &  969 + 49 sec                      &  12 sec  \\
\hline
Lane 6, 10 samples  &  924 + 5133 sec       &  924 + 186 sec                     &  16 sec  \\
\hline
Whole dataset & $2.3\,10^5$ sec (est.) & $5.5\, 10^4$ sec (est.)    &  626 sec (actual)  \\
\hline
\end{tabular}
\caption{Runtime comparison of DADA and the simplified cluster-free filtering (CFF) denoiser on two representative sets of 10 tongue samples (lane 5 and lane 6). Lanes were considered separately since samples in the two groups tended to differ significantly in the number of reads retained by quality filtering. The whole dataset consisted of 189 samples on lane 5 and 320 samples on lane 6. Comparisons were performed on an Intel Xeon CPU 2.83GHz.}
\label{tbl:DADAtime}
\end{table}

\subsection{Example of other applications: environmental cross-sectional 454 data}
The approach described in this work does not explicitly rely on the longitudinal nature of the sampling. Most of our analysis can be readily applied to any multi-sample datasets, e.g. a cross-sectional sampling or a location series, provided samples were collected and processed in a similar way so that the error structure can be assumed to be similar. Further, and despite the caveats we described, our method can be applied even to data collected using the 454 sequencing platform. To illustrate the broad applicability of our approach, we used data from a cross-sectional environmental sampling conducted by Preheim et al. (SRA accession number from SRP029470). Lake water microbiota were sampled at depths ranging from 0 m (surface) to 22 m with 1-meter depth intervals. The authors used this data to illustrate their sequence clustering algorithm (DBC) that also relies on cross-sample comparisons to distinguish between closely related OTUs; for details, see the original reference (Preheim et al., 2013). They report their algorithm worked best with stringent quality filtering whereupon sequences were trimmed to just 76 nt, and any reads containing bases with Phred quality scores at or below 16 were discarded. This filtering retained 7.78M total sequences (120K unique). Since our approach includes data denoising, we could use much more liberal quality score filtering and retain more data (USEARCH maxEE of 1 and truncating at Phred quality score 2). To compare runtime of our algorithm and DBC, we increased the read truncation length so as to keep the same total number of sequences. This set the quality-filtered sequence length to $L=91$ nt, 20\% longer than used by the authors (7.98M sequences, 300K unique; \verb|tcsh run_CFF_on_FastQ.tcsh 91 analysisResults "PreheimData/*.fastq"|.).

\begin{figure}[b]
\includegraphics[width = 0.95 \textwidth]{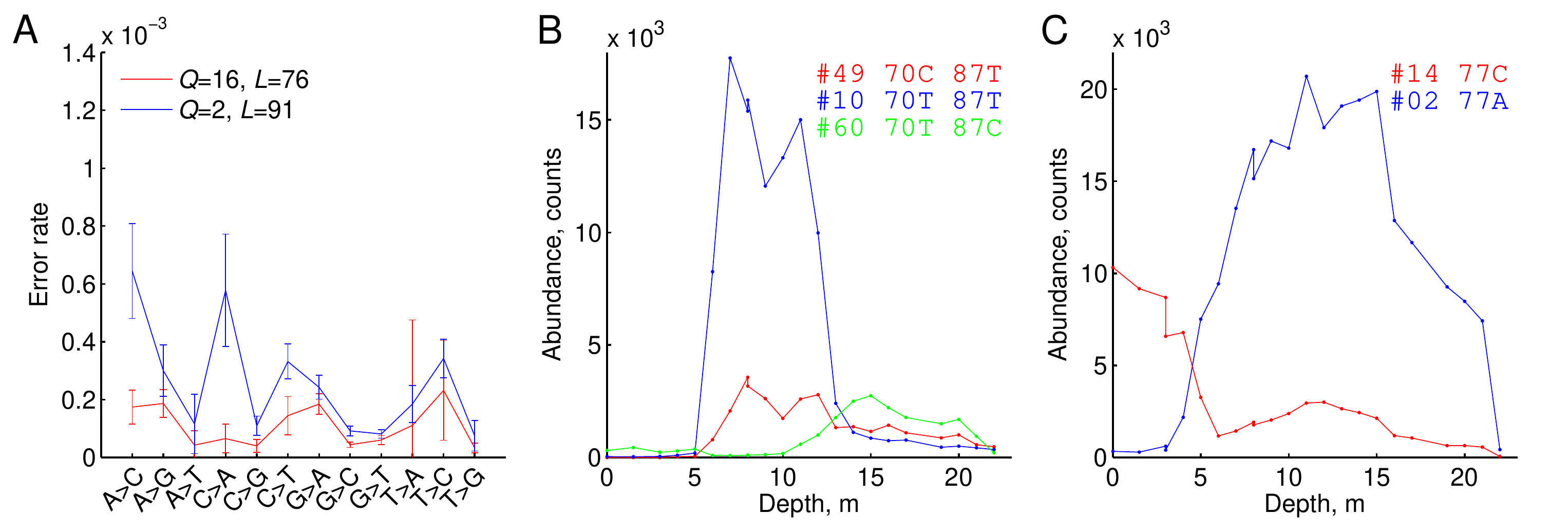}
\caption{Cross-sectional environmental 454 data: lake water microbiota (Lake Mystic) sampled at depths 0--22m. \textbf{A:} Substitution error rates inferred from the data for two sets of quality filtering parameters indicated in the legend; $Q$ is the Phred quality score truncation threshold; $L$ is read truncation length. \textbf{B:} Three sequences resolved by our analysis; Seq.~\#49 and Seq.~\#60 both differ from Seq.~\#10 by a single nucleotide at locations 70 and 87, respectively. Sequence abundance is shown as a function of depth; sequences are labeled by cumulative abundance rank. \textbf{C:} Same, for sequences Seq.~\#2 and Seq.~\#14 differing at nucleotide 77.}
\label{fig:DBC}
\end{figure}
Fig.~\ref{fig:DBC}A shows the substitution error rates inferred from the data at both sets of quality filtering parameters. Note that these rates are significantly lower than those of Fig.~\ref{fig:QCparams} (the scales of the two plots are identical), exhibit a very weak transition/transversion bias, and are more sensitive to Phred score quality filtering than what we have seen with Caporaso et al. data (Fig.~\ref{fig:QCparams}). This seems to indicate that the protocol used by Preheim et al. generates significantly fewer PCR substitution errors. This  dependence of error rates on the experimental protocol highlights the advantage of being able to estimate error rates for a given dataset directly from the data, without the need for a separate calibration.

Fig.~\ref{fig:DBC}BC provide examples of sequences differing by a single nucleotide exhibiting ecologically significant distinctions, as identified by our method in this environmental dataset; compare with Fig.~2A, Fig.~\ref{fig:BirthDeath} and Fig.~\ref{fig:Feces}AB. Sequence abundance is shown as a function of depth (each sample was independently normalized to $3.2\,10^5$ total quality-filtered reads per sample, to correct for varying sample size). The DBC method of Preheim et al. is also capable of identifying OTUs differing by a single nucleotide (compare Fig.~\ref{fig:DBC}BC to Fig.~5b in the original reference); however, our analysis achieved higher resolution by retaining longer reads and took only 13 min single-core processor time (Intel Xeon CPU 2.83GHz), compared to 8 hours analysis time the authors report for parallelized DBC running on a cluster with 60-100 processes executing simultaneously. In fact, the true runtime difference is even greater, since the complexity of both algorithms scales with the number of \textit{unique} sequences rather than total reads. For $Q_\mathrm{min}=17, L=76$ as used by the authors of DBC, our algorithm completes in only 210 seconds.

We stress, however, that DBC and cluster-free filtering seek to achieve different goals and are not directly comparable. DBC is an OTU clustering algorithm, whereas the goal of cluster-free filtering is to identify sub-OTU structure of moderate-to-high-abundance community members. However, to our knowledge DBC is the only existing tool that exploits cross-sample comparisons to inform the interpretation of sequencing data, and the performance comparison above serves to illustrate the drastically different computational cost of the two approaches.

\subsection{How many samples is enough?}\label{sec:numberOfSamples}
We have described a method that employs cross-sample comparisons to achieve sub-OTU resolution. The analysis presented in the main text uses data from a study with an uncommonly large number of samples; in contrast, the previous section demonstrates that our method can be usefully applied to a dataset with only 22 datapoints. What is the minimum number of samples required by our method?

The answer is that the number of samples determines the resolution that can be achieved; more samples will always allow higher resolution, but coarser differences can be resolved with just a few. For example, just 2 samples (say, 0m and 10m) would have been enough to resolve the two subpopulations presented on Fig.~\ref{fig:DBC}C. By contrast, the difference between depth traces of Seq.~\#10 and Seq.~\#49 (Fig.~\ref{fig:DBC}B) is less pronounced and more samples are required. Finally, resolving the sequences in Fig.~2B would not have been possible with fewer than $\approx$100 samples. For high-abundance sequences where the complex struture of noise in the counts can be neglected, this tradeoff can be formally quantified using the Jensen-Shannon divergence as a measure of distance between abundance distributions of two sequences across samples; for details, see Preheim et al., 2013.

\newpage
\section{Supplementary information for Figure 2}
\subsection{A pair of sequences representing strongly anticorrelated subpopulations}
\begin{figure}[h!]
\includegraphics[width = 0.65 \textwidth]{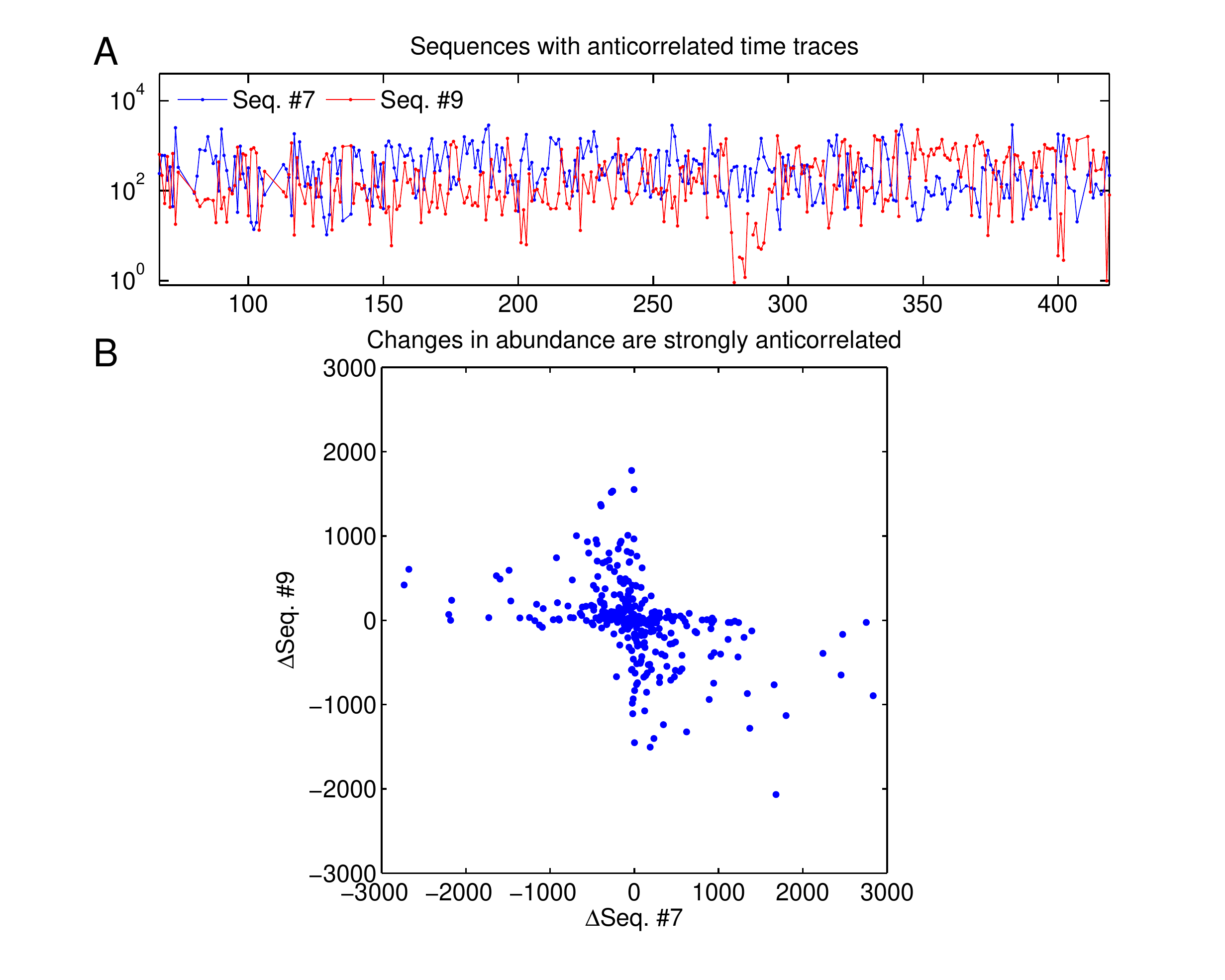}
\caption{\textbf{A.} Abundance time traces of Seq.~\#7 and Seq.~\#9. \textbf{B.} Scatter plot for the same sequences of their discrete derivatives of abundance (i.e. abundance changes from each day to the next). A BLAST search against the GreenGenes database identifies the likely taxonomy of Seq.~\#7 as \textit{Streptococcus thermophilus}. Seq.~\#9 does not have a good match; the closest hit is an unclassified \textit{Prevotella sp.} at only 88\% identity.}
\label{fig:Anticor}
\end{figure}

\subsection{Best expected correlation of two time traces}
The maximum degree to which time traces of two sequences can be correlated is a function of their abundance: for low-abundance sequences the Poisson sampling noise becomes non-negligible and sets an upper bound for the best achievable correlation coefficient. Consequently, to define a correlation as strong or weak, any measured correlation coefficient should be compared to this abundance-dependent quantity rather than to 1.

Let $N(t)$ be the true abundance time trace of some bacterial strain (in units of cells, rather than sequence counts). Imagine that two sequences in the dataset were measuring the abundance of this exact same strain, but with different amplification efficiencies $\lambda_1$ and $\lambda_2$ (let $\lambda_1>\lambda_2$). Neglecting all sources of noise other than the Poisson counting noise, the abundance traces of these two sequences can be modeled by

$$
n_{1,2} (t)=\Poiss [\lambda_{1,2} N(t)],
$$
where $\Poiss[\cdot]$ denotes adding Poisson noise. Since Poisson noise is unavoidable, the correlation coefficient between these two traces sets an upper bound for the correlation between $n_1 (t)$ and any other trace $n^\ast (t)$ with the same mean abundance as $n_2 (t)$. This maximum correlation depends on the shape of the trace $N(t)$ and amplification efficiencies $\lambda_1$, $\lambda_2$, and can be expressed as follows:

$$
c_{\mathrm{max}}[N(t),\lambda_1,\lambda_2]=\corr\left(\Poiss[\lambda_1 N(t)],\Poiss[\lambda_2/\lambda_1 \ast \lambda_1 N(t)]\right).
$$

And therefore, in terms of measurable quantities only:

$$
c_{\mathrm{max}}[n_1(t),\langle n_2\rangle] \approx \corr\left(\Poiss[n_1(t)],\Poiss[\langle n_2\rangle/\langle n_1\rangle \ast n_1(t)]\right).
$$

Here $\langle\cdot\rangle$ denotes the average abundance, and we use the higher-abundance trace of the pair as the best estimate of the shape of the true abundance $N(t)$. The maximum correlation coefficient depends on the shape of the trace $n_1 (t)$ and on the mean abundance of the trace we compare it to; the lower the mean abundance is, the stronger the effect of Poisson noise and the lower the $c_{\mathrm{max}}$.

In practice, for a pair of traces $n_1 (t)$, $n_2 (t)$, we compute their best expected correlation as follows:
\begin{enumerate}
\item	Take the more abundant trace $n_1 (t)$
\item	Construct a renormalized trace $n_2^{\mathrm{mock}}(t)=\frac{\langle n_2\rangle}{\langle n_1\rangle} n_1 (t)$
\item	Poisson-resample both of these 10 times: denote these $n_1^{(i)}$, $n_2^{\mathrm{mock}(i)}$, $i=1\dots 10$.
\item	Compute the 100 correlation coefficients between all pairs $c_{ij}=\corr\left(n_1^{(i)},n_2^{\mathrm{mock}(j)}\right)$.
\item	Set $c_{\mathrm{max}}[n_1 (t),\langle n_2\rangle]=\langle c_{ij}\rangle$.
\end{enumerate}

\begin{figure}[t!]
\includegraphics[width = 0.4 \textwidth]{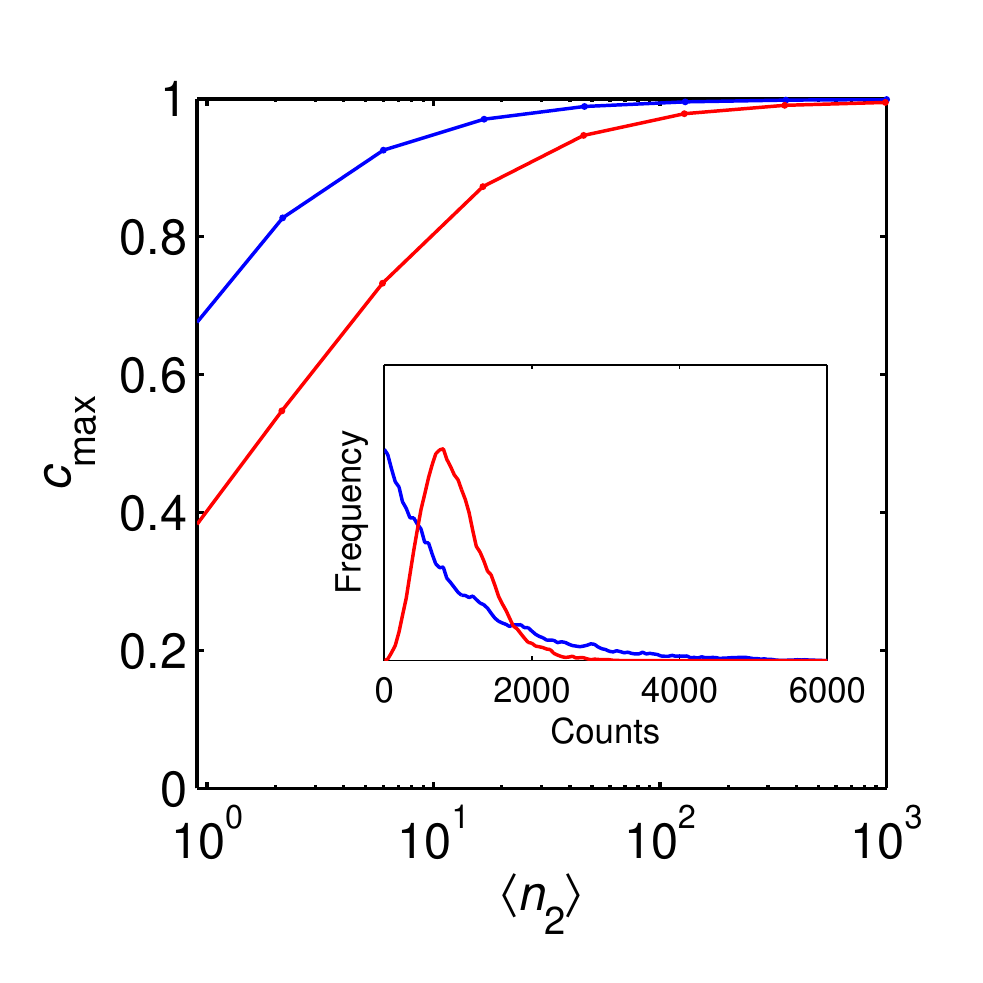}
\caption{Best expected correlation $c_{\mathrm{max}}$  for a pair of abundance time traces $n_{1,2} (t)$ is sensitive to the shape of the distribution of the daily counts, not just the average abundance $\langle n_{1,2}(t)\rangle$. The figure shows the best expected correlation $c_{\mathrm{max}} [n_1 (t),\langle n_2 \rangle]$ as defined in the Supplementary Information, for two different mock traces $n_1 (t)$ with the same mean ($\langle n_1\rangle=1000\text{ counts}/\mathrm{day}$) but with different distributions, modeled here by Gamma distributions with shape parameter 1 (blue) and 5 (red). The best expected correlation increases with the mean abundance $\langle n_2 (t)\rangle$, but for the same mean it is higher for the blue trace whose distribution covers a wider dynamic range. Because of this nontrivial dependence on the distribution shape, in our definition of dynamical similarity we compute the best expected correlation individually for every pair of sequences.}
\label{fig:MaxCorr}
\end{figure}

The shape of the function $c_{\mathrm{max}}[n_1 (t),\langle n_2\rangle]$ is illustrated on Fig.~\ref{fig:MaxCorr}.

\subsection{Distance metric for sequence pairs}
We use the BLAST definition, i.e. the ratio of the number of mismatches to the total number of columns after pairwise realignment, and multiply this ratio by the length of the sequence (130 nt). For close sequences that differ by a few substitution errors the alignment is trivial, and this normalization corresponds to the Hamming distance between sequences, in nt.

\newpage
\section{Supplementary information for Figure 3}
\subsection{Estimating correlation time from autocorrelation function}
We define the autocorrelation time $\tau$ of a sequence as the time shift $\Delta t$ at which the autocorrelation function $c_{\Delta t}$ falls below the threshold of statistical significance. For reasons discussed above, the notions of strong (significant) or weak (insignificant) correlation of sequence time traces are abundance-dependent. Therefore, instead of using a fixed threshold value for all sequences, we proceed in the following way. For a given sequence, we first compute its root-mean-square autocorrelation coefficient for time shifts between 70 and 100 samples:

$$
c_{\mathrm{null}} =\sqrt{\langle(c_{\Delta t})^2\rangle_{\Delta t=70\dots100}}.
$$

If we assume that all autocorrelation observed at such large time shifts is entirely due to noise, then $c_{\mathrm{null}}$  provides a natural scale for statistical significance. We conservatively define the significance threshold at twice the magnitude of $c_{\mathrm{null}}$.

Note that $c_{\mathrm{null}}$ provides an upper bound on a statistically significant correlation value. If some dynamical processes in the population are slow enough that they contribute to the autocorrelation function even at such large time shifts (\textit{cf.} Fig.~\ref{fig:BirthDeath}), this will increase $c_{\mathrm{null}}$ and cause us to underestimate the true autocorrelation time. This means that assuming $c_{\mathrm{null}}$ was entirely due to noise is a safe approximation to make: if it does not hold, it can only strengthen our conclusion that the sequence abundance time traces exhibit multi-day autocorrelations.

\subsection{Examples of sequences exhibiting consistent dynamics on very long time scales}
\begin{figure}[h!]
\includegraphics[width = 0.65 \textwidth]{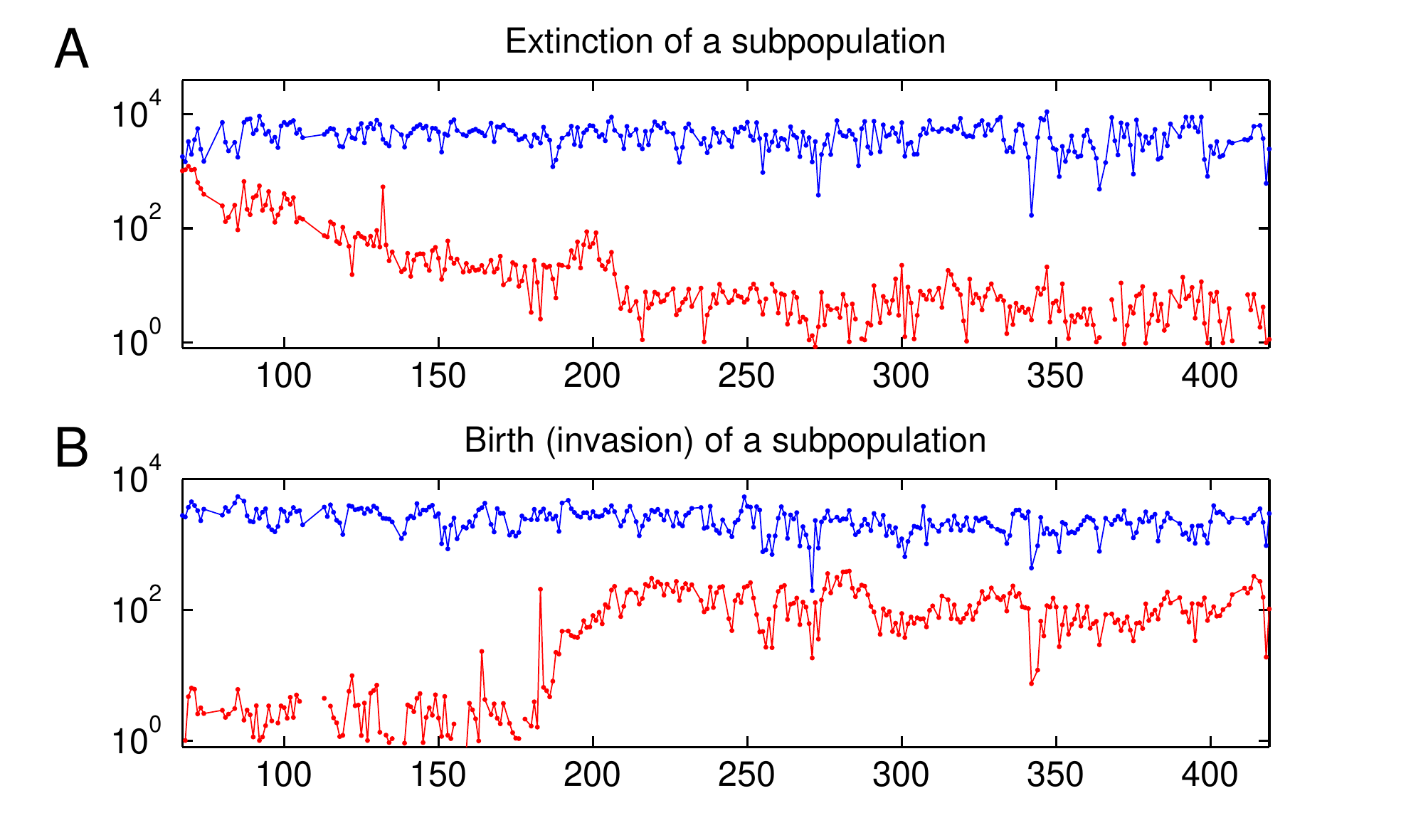}
\caption{\textbf{A.} Slow extinction of a subpopulation (red; cf. Fig.~1B, neighbor 3). From day 210 onwards the abundance of the red sequence is consistent with it being a substitution error of Seq.~\#1 (blue), which is a direct neighbor in sequence space. \textbf{B.} Slow birth/invasion of a subpopulation (red). The new sequence differs by 1nt from well-established Seq.~\#2 (blue), and prior to day 160 its abundance is consistent with being its substitution error. Note the high similarity of fluctuations from day 210 onwards.}
\label{fig:BirthDeath}
\end{figure}

\begin{figure}[h!]
\includegraphics[width = 0.65 \textwidth]{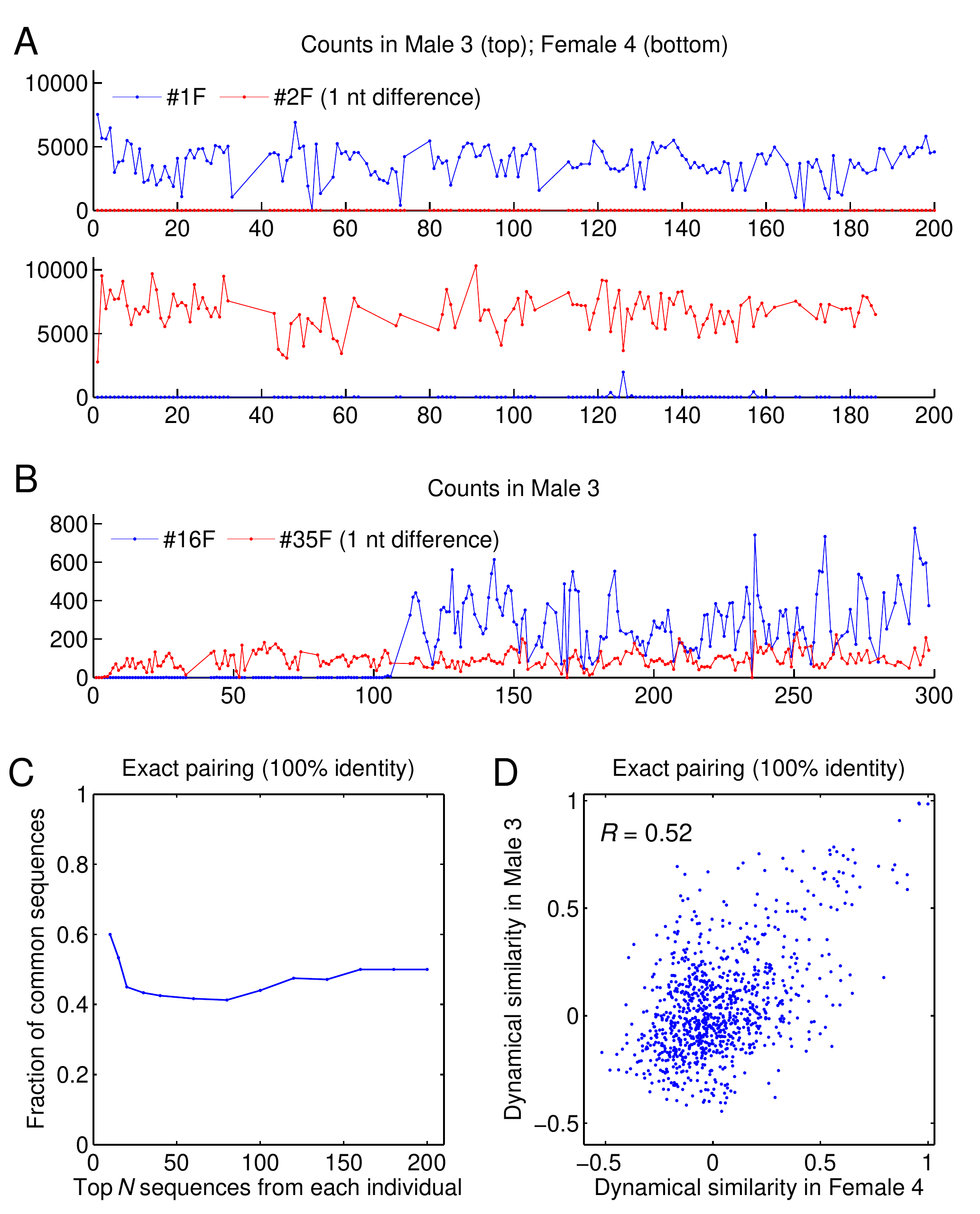}
\caption{\textbf{A.} Abundance time traces (sequence counts \emph{vs.} observation day) for Seq.~\#1F and Seq.~\#2F, which differ by a single nucleotide and dominate in individuals Male 3 and Female 4, respectively. \textbf{B.} Another example of abundance time traces of two sequences that differ by a single nucleotide (99.2\% similarity), yet exhibit strongly distinct dynamics and so derive from distinct bacteria. \textbf{C.} Fraction of shared 16S sequences, defined as the fraction of common tags (at 100\% sequence identity) among the most abundant $N$ sequences in the fecal samples of each of the two individuals, plotted as a function of $N$ (compare with Fig.~5A.) \textbf{D.} Scatter plot of the dynamical similarity of pairs of common fecal sequences, as measured independently in the two individuals, for all possible pairs among the 44 common sequences shared within the top $N = 100$ (compare with Fig.~5B).}
\label{fig:Feces}
\end{figure}

Fig.~\ref{fig:BirthDeath}AB shows examples of sequences exhibiting steady change in abundance for more than a month. In both cases, the slow-changing sequence is 99.2\% similar to a very high-abundance community member and could not have been resolved by traditional OTU-based methods. Note the sharp jump in panel B at day 182 of the sequence representing the invading subpopulation (red) to an abundance value close to the equilibrium established after day 210. It is intriguing to speculate that this trace may document spatial invasion of a subpopulation already established elsewhere on the tongue, a region accidentally sampled on day 182.

\subsection{Persistence of difference: the null model}
To distinguish between 16S tags coming from distinct subpopulations or from physically the same bacterial cells, we introduced a quantity we called the \emph{persistence of difference} $P_D$. For this, we first defined the fractional difference $\Delta(t)$ between two time traces renormalized to the same mean $n_{A,B} (t)$:

$$
\Delta(t)=\frac{n_A-n_B}{(n_A+n_B)/2}.
$$

We then defined the persistence of difference $P_D$ as the 1-day autocorrelation coefficient of $\Delta(t)$. If $A$ and $B$ are two genomic variants contained within the same bacterium, then any difference between $n_A (t)$ and $n_B (t)$ must be due to measurement noise, and $P_D$ must vanish. If, however, $n_{A,B} (t)$ reflect abundances of two distinct subpopulations, then $\Delta(t)$ can be expected to exhibit some degree of autocorrelation due to the slow dynamics observed for most individual sequences. We gave an intuitive argument for this in the main text. Here, to gain some extra intuition about the null model for $P_D$, we calculate it explicitly in the simplest case when the two traces $n_{A,B}(t)$ are independent and can be approximated by a stationary, weakly fluctuating process:

\begin{align}
n_A (t)&=\mu\left(1+\sigma_A \xi_A (t)\right)\\
n_B (t)&=\mu\left(1+\sigma_B \xi_B (t)\right)
\end{align}

Here $\xi_{A,B}$ have zero mean, unit variance and are uncorrelated. Assuming $\sigma_{A,B}\ll1$, we can write:

$$
\Delta(t)\approx \sigma_A \xi_A (t)-\sigma_B \xi_B (t)
$$

And therefore, making use of the independence assumption,

$$
P_D=\frac{\langle\Delta(t)\Delta(t+1)\rangle}{\langle\Delta(t)^2\rangle}
\approx\frac{\langle\sigma_A^2\xi_A(t)\xi_A(t+1)+\sigma_B^2\xi_B(t)\xi_B(t+1)\rangle}{\sigma_A^2+\sigma_B^2}
=\frac{\sigma_A^2c_{1A}+\sigma_B^2c_{1B}}{\sigma_A^2+\sigma_B^2}
$$

Here $c_{1A,B}$ are the one-day autocorrelation coefficients of the fluctuations of the two individual sequences.

The independence approximation made above is clearly not valid for the dynamics of most community members. For this reason, for the purposes of Fig.~3C, the null-model prediction was constructed directly from the data, by reversing in all pairs the time order for one of the sequences prior to the calculation of $P_D$. This removes any real correlations of the traces while preserving autocorrelation and other properties of the traces such as their fluctuation spectrum. Nevertheless, the calculation above is useful as it explains why the null-model expectation for $P_D$ is non-zero when both sequences have slow internal dynamics.

Note that a sequence with an exceptionally long intrinsic time scale (as shown in Fig.~\ref{fig:BirthDeath}AB) will have a large $P_D$ score when paired with any other sequence. These two sequences were therefore excluded from Fig.~3C.

\subsection{Persistence of difference for non-longitudinal data}
None of the cross-sample comparison methodology described in this work is limited to time series data. The ``persistence of difference'' argument accompanying Fig.~4 is no exception; however, it does rely on two additional assumptions, namely that the composition of samples varies smoothly with some parameter labeling the samples, and that the sampling frequency is sufficiently high to allow correlations of fluctuations to be observed between consecutive samples. For the longitudinal data series of Caporaso et al. this parameter was time; for a location series one can expect community composition to vary smoothly in space, and the same argument can be applied. In other words, the use of ``persistence of difference'' $P_D$ need not be limited strictly to longitudinal datasets. However, autocorrelation-based analysis is particularly sensitive to the number of samples (see section~\ref{sec:numberOfSamples}). Determining whether $P_D$ can be a useful concept for studying the spatial heterogeneity of populations requires further investigation.

\newpage
\section{Supplementary information for Figure 4}
\subsection{Over-estimation of OTU quality scores}
As described in the main text, for the purposes of Fig.~4, when calculating OTU quality scores, we restricted our attention only to high-abundance members of the OTU, considering only sequences from the top 200 by overall abundance. Since most of the diversity is contributed by low-abundance species (Huttenhower et al., 2012), Fig.~4 underestimates the true diversity of an OTU. Including lower-abundance OTU members makes OTU quality scores drop continuously as new OTU members are added; however, it also becomes increasingly hard to separate dynamical diversity from the effects of noise. Consequently, in Fig.~4 we report our most conservative estimate of within-OTU diversity, where we use only the highest-abundance members out of all those resolved by cluster-free filtering (there was an average of $18\pm4$ resolved sequences within a 97\% OTU, and only $9\pm2$ per OTU were used for Fig.~4).

In addition, OTU quality scores were calculated under the assumption that each sequence represents a separate subpopulation. Sequences that in fact derive from the same bacteria (16S paralogs or errors not in our model) appear in the defining equation as independent, dynamically identical subpopulations, increasing the apparent OTU quality score. This is another reason why the true quality scores of OTUs are likely even lower than reported in Fig.~4.

\newpage
\section{Supplementary information for Figure 5}
\subsection{Cross-individual analysis of fecal samples}
To confirm our conclusions from the analysis of tongue microbiome data presented in the main text, we repeated our analysis using fecal samples of the two individuals, collected in the same study (Caporaso et al., 2011). There were 374 samples, 243 from the male subject and 131 from the female, with $2.5\pm0.5\, 10^4$ reads per sample. We normalized the observed abundances to $2.5\,10^4$ total reads in each sample to correct for varying sample size. As before, we labeled sequences in order of decreasing overall abundance (pooling samples from both individuals): Seq~\#1F, Seq~\#2F, etc., where ``F'' reflects that we are now dealing with fecal samples rather than the tongue.

Again, we find that sequences differing by as little as a single nucleotide can exhibit ecologically significant differences in their dynamics. The most striking example is that the dominating sequence in individual ``Male 3'' differs from the dominating sequence in ``Female 4'' by a single nucleotide, and virtually no cross-contamination is observed (Fig/~\ref{fig:OTU}A). Both these sequences map to \textit{Bacteroides sp.} in GreenGenes (DeSantis et al., 2006). Another example is presented in panel B. Finally, we repeat the cross-individual analysis presented, for tongue samples, in Fig.~5AB. We find that the two gut communities as probed by the fecal samples also share a large fraction of sequences at 100\% identity. This once again supports the scenario whereby the communities exchange members with non-negligible frequency, although less so than the tongue samples. The observation of panel A is therefore unlikely to represent the effect of dispersal limitation, suggesting instead a functional difference between the representatives of \textit{Bacteroides sp.} established in the two individuals or a resistance to invasion. Finally, we find that the dynamical similarity of shared sequences, when measured independently in the two individuals, is clearly correlated, just as it was for the tongue communities (Fig.~5B).  With the number of shared sequences being lower for fecal samples than for tongue samples, the statistics were insufficient to compare dynamical similarity of ``intentionally mismatched'' sequences as in Fig.~5C.

\subsection{Cross-individual analysis at 97\% OTU level}
\begin{figure}[h!]
\includegraphics[width = 0.5 \textwidth]{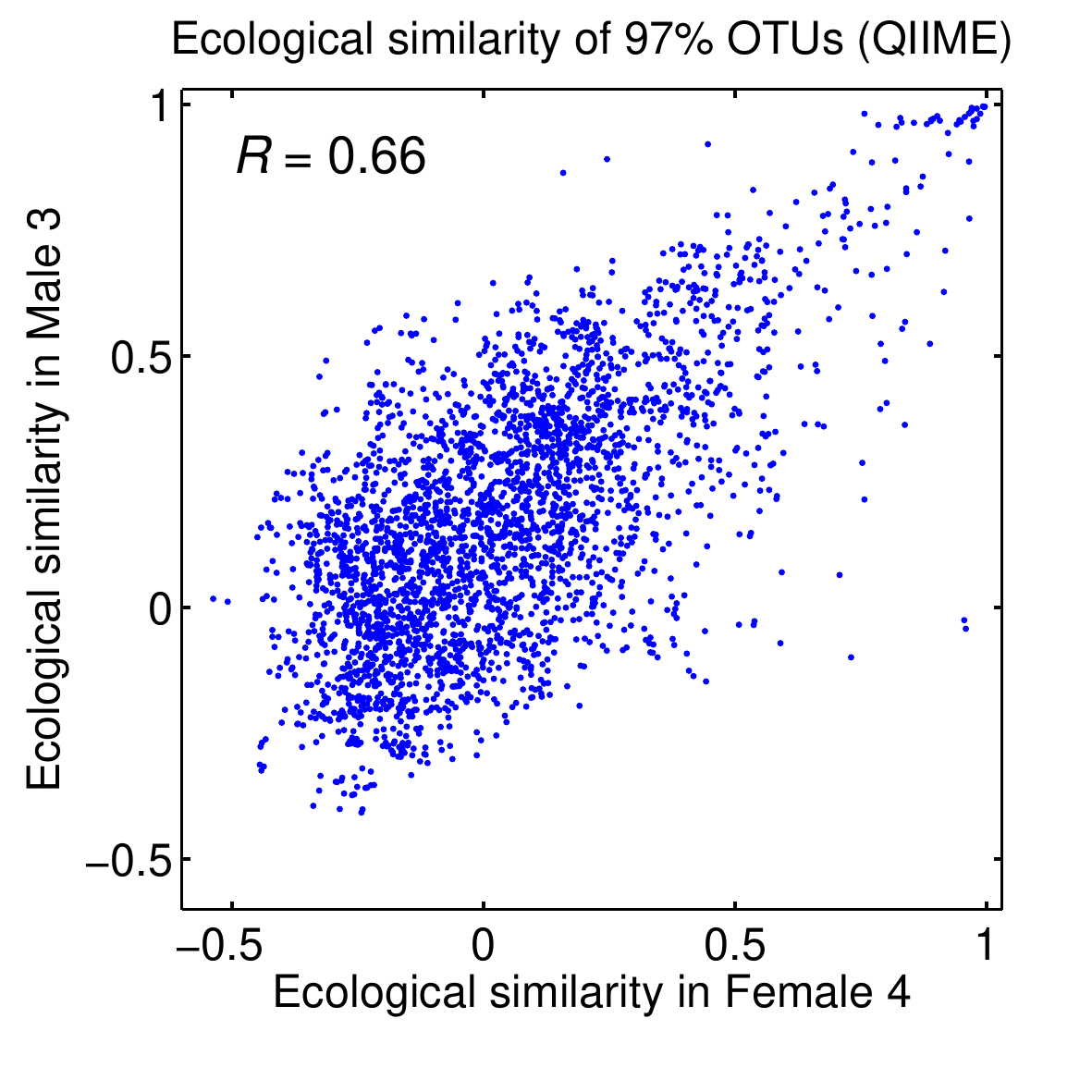}
\caption{Dynamical similarity between pairs of common 97\% OTUs, as measured independently in the two individuals, for 78 common OTUs within the top 100, constructed using closed-reference OTU picking as implemented in QIIME.}
\label{fig:OTU}
\end{figure}
The same analysis as in Fig.~5B can be performed for shared 97\% OTUs rather than shared sequences (at 100\% identity). We constructed OTUs using closed-reference OTU picking as implemented in QIIME, matching sequences at 97\% sequence similarity against the GreenGenes database. Fig.~\ref{fig:OTU} shows the scatter plot of the dynamical similarity between pairs of common OTUs, as measured independently in the two individuals, for 78 common OTUs (those shared within the top 100). Note, however, that most OTUs are dominated by a single high-abundance sequence (as evidenced by the high weighted quality score on Fig.~4), and most of these dominating sequences are shared across the two communities (Fig.~5A). For these reasons, the plot shown here is very similar to Fig.~5B, but only because the within-OTU diversity is masked by dominating subpopulations.

\newpage
\section*{Supplementary references}
\addcontentsline{toc}{section}{Supplementary references}
\leftskip 0.2in
\parindent -0.2in
\parskip 1em


DeSantis TZ, Hugenholtz P, Larsen N, Rojas M, Brodie EL, Keller K et al (2006). Greengenes, a chimera-checked 16S rRNA gene database and workbench compatible with ARB. {\em Appl Environ Microbiol} {\bf 72}: 5069--5072.

Edgar RC (2004). MUSCLE: multiple sequence alignment with high accuracy and high throughput. {\em Nucleic Acids Res.} {\bf 32}:1792--1797.



\begin{thebibliography}{99\kern\bibindent}
\makeatletter
\def\@biblabel#1{}
\let\old@bibitem\bibitem
\def\bibitem#1{\old@bibitem{#1}\leavevmode\kern-\bibindent}
\makeatother

\bibitem{Brestoff13}
Brestoff JR, Artis D (2013). Commensal bacteria at the interface of host metabolism and the immune system. {\em Nat Immunol} {\bf 14}: 676-684.
%
\bibitem{Caporaso11}
Caporaso JG, Lauber CL, Costello EK, Berg-Lyons D, Gonzalez A, Stombaugh J et al (2011). Moving pictures of the human microbiome. {\em Genome biology} {\bf 12}: R50.
%
\bibitem{Caporaso12}
Caporaso JG, Lauber CL, Walters WA, Berg-Lyons D, Huntley J, Fierer N et al (2012). Ultra-high-throughput microbial community analysis on the Illumina HiSeq and MiSeq platforms. {\em ISME J} {\bf 6}: 1621-1624.
%
\bibitem{Costello09}
Costello EK, Lauber CL, Hamady M, Fierer N, Gordon JI, Knight R (2009). Bacterial community variation in human body habitats across space and time. {\em Science} {\bf 326}: 1694-1697.
%
\bibitem{Costello12}
Costello EK, Stagaman K, Dethlefsen L, Bohannan BJM, Relman DA (2012). The application of ecological theory toward an understanding of the human microbiome. {\em Science} {\bf 336}: 1255-1262.
%
\bibitem{DeSantis06}
DeSantis TZ, Hugenholtz P, Larsen N, Rojas M, Brodie EL, Keller K et al (2006). Greengenes, a chimera-checked 16S rRNA gene database and workbench compatible with ARB. {\em Appl Environ Microbiol} {\bf 72}: 5069-5072.
%
\bibitem{Edgar04}
Edgar RC (2004). MUSCLE: multiple sequence alignment with high accuracy and high throughput. {\em Nucleic Acids Res.} {\bf 32}: 1792-1797
%
\bibitem{Edgar10}
Edgar RC (2010). Search and clustering orders of magnitude faster than BLAST. {\em Bioinformatics} {\bf 26}: 2460-2461.
%
\bibitem{Edgar11}
Edgar RC, Haas BJ, Clemente JC, Quince C, Knight R (2011). UCHIME improves sensitivity and speed of chimera detection. {\em Bioinformatics} {\bf 27}: 2194-2200.
%
\bibitem{Eren13}
Eren AM, Maignien L, Sul WJ, Murphy LG, Grim SL, Morrison HG et al (2013). Oligotyping: differentiating between closely related microbial taxa using 16S rRNA gene data. {\em Methods in Ecology and Evolution} {\bf 4}: 1111--1119.
%
\bibitem{Faith13}
Faith JJ, Guruge JL, Charbonneau M, Subramanian S, Seedorf H, Goodman AL et al (2013). The long-term stability of the human gut microbiota. {\em Science} {\bf 341}: 1237439.
%
\bibitem{Fierer11}
Fierer N, Lennon JT (2011). The generation and maintenance of diversity in microbial communities. {\em Am J Bot} {\bf 98}: 439-448.
%
\bibitem{Fredricks13}
Fredricks DN (2013). {\em The human microbiota: how microbial communities affect health and disease.} Wiley-Blackwell: Hoboken, N.J.
%
\bibitem{Haas11}
Haas BJ, Gevers D, Earl AM, Feldgarden M, Ward DV, Giannoukos G et al (2011). Chimeric 16S rRNA sequence formation and detection in Sanger and 454-pyrosequenced PCR amplicons. {\em Genome Res} {\bf 3}: 494-504.
%
\bibitem{Hamady09}
Hamady M, Knight R (2009). Microbial community profiling for human microbiome projects: Tools, techniques, and challenges. {\em Genome Res} {\bf 19}: 1141-1152.
%
\bibitem{Huang10}
Huang Y, Niu BF, Gao Y, Fu LM, Li WZ (2010). CD-HIT Suite: a web server for clustering and comparing biological sequences. {\em Bioinformatics} {\bf 26}: 680-682.
%
\bibitem{Hunt08}
Hunt DE, David LA, Gevers D, Preheim SP, Alm EJ, Polz MF (2008). Resource partitioning and sympatric differentiation among closely related bacterioplankton. {\em Science} {\bf 320}: 1081-1085.
%
\bibitem{Huse10}
Huse SM, Welch DM, Morrison HG, Sogin ML (2010). Ironing out the wrinkles in the rare biosphere through improved OTU clustering. {\em Environ Microbiol} {\bf 12}: 1889-1898.
%
\bibitem{Huttenhower12}
Huttenhower C, Gevers D, Knight R, Abubucker S, Badger JH, Chinwalla AT et al (2012). Structure, function and diversity of the healthy human microbiome. {\em Nature} {\bf 486}: 207-214.
%
\bibitem{Kamada13}
Kamada N, Chen GY, Inohara N, Nunez G (2013). Control of pathogens and pathobionts by the gut microbiota. {\em Nat Immunol} {\bf 14}: 685-690.
%
\bibitem{Klindworth13}
Klindworth A, Pruesse E, Schweer T, Peplies J, Quast C, Horn M et al (2013). Evaluation of general 16S ribosomal RNA gene PCR primers for classical and next-generation sequencing-based diversity studies. {\em Nucleic Acids Res} {\bf 41}.
%
\bibitem{Kunin10}
Kunin V, Engelbrektson A, Ochman H, Hugenholtz P (2010). Wrinkles in the rare biosphere: pyrosequencing errors can lead to artificial inflation of diversity estimates. {\em Environ Microbiol} {\bf 12}: 118-123.
%
\bibitem{Langille13}
Langille MGI, Zaneveld J, Caporaso JG, McDonald D, Knights D, Reyes JA et al. (2013) Predictive functional profiling of microbial communities using 16S rRNA marker gene sequences. {\em Nature Biotechnol} {\bf 31}: 814–821.
%
\bibitem{Lozupone05}
Lozupone C, Knight R (2005). UniFrac: a new phylogenetic method for comparing microbial communities. {\em Appl Environ Microbiol} {\bf 71}: 8228-8235.
%
\bibitem{Lukjancenko10}
Lukjancenko O, Wassenaar TM, Ussery DW (2010). Comparison of 61 sequenced Escherichia coli genomes. {\em Microbial ecology} {\bf 60}: 708-720.
%
\bibitem{Morgan13}
Morgan MJ, Chariton AA, Hartley DM, Court LN, Hardy CM (2013). Improved inference of taxonomic richness from environmental DNA. {\em Plos One} {\bf 8}: e71974.
%
\bibitem{Ochman03}
Ochman H (2003). Neutral mutations and neutral substitutions in bacterial genomes. {\em Mol Biol Evol} {\bf 20}: 2091-2096.
%
\bibitem{Preheim13}
Preheim SP, Perrotta AR, Martin-Platero AM, Gupta A, Alm EJ (2013). Distribution-Based Clustering: Using Ecology To Refine the Operational Taxonomic Unit. {\em Appl Environ Microbiol} {\bf 79}: 6593-6603.
%
\bibitem{Prosser07}
Prosser JI, Bohannan BJM, Curtis TP, Ellis RJ, Firestone MK, Freckleton RP et al (2007). Essay --- The role of ecological theory in microbial ecology. {\em Nature Reviews Microbiology} {\bf 5}: 384-392.
%
\bibitem{Quince09}
Quince C, Lanzen A, Curtis TP, Davenport RJ, Hall N, Head IM et al (2009). Accurate determination of microbial diversity from 454 pyrosequencing data. {\em Nat Methods} {\bf 6}: 639-U627.
%
\bibitem{Quince11}
Quince C, Lanzen A, Davenport RJ, Turnbaugh PJ (2011). Removing noise from pyrosequenced amplicons. {\em BMC Bioinformatics} {\bf 12}: 38.
%
\bibitem{Rosen12}
Rosen MJ, Callahan BJ, Fisher DS, Holmes SP (2012). Denoising PCR-amplified metagenome data. {\em BMC Bioinformatics} {\bf 13}: 283.
%
\bibitem{Schloss09}
Schloss PD, Westcott SL, Ryabin T, Hall JR, Hartmann M, Hollister EB et al (2009). Introducing mothur: Open-Source, Platform-Independent, Community-Supported Software for Describing and Comparing Microbial Communities. {\em Appl Environ Microbiol} {\bf 75}: 7537-7541.
%
\bibitem{Schloss11}
Schloss PD, Gevers D, Westcott SL (2011). Reducing the effects of PCR amplification and sequencing artifacts on 16S rRNA-based studies. {\em PLOS One} {\bf 6}.
%
\bibitem{SchlossWestcott11}
Schloss PD, Westcott SL (2011). Assessing and improving methods used in operational taxonomic unit-based approaches for 16S rRNA gene sequence analysis. {\em Appl Environ Microbiol} {\bf 77}: 3219-3226.
%
\bibitem{Shade12}
Shade A, Peter H, Allison SD, Baho D, Berga M, Buergmann H et al (2012). Fundamentals of microbial community resistance and resilience. {\em Frontiers in Microbiology} {\bf 3}.
%
\bibitem{Shade13}
Shade A, Caporaso JG, Handelsman J, Knight R, Fierer N (2013). A meta-analysis of changes in bacterial and archaeal communities with time. {\em ISME J} {\bf 7}: 1493-1506.
%
%
\bibitem{Song13}
Song SJ, Lauber C, Costello EK, Lozupone CA, Humphrey G, Berg-Lyons D et al (2013). Cohabiting family members share microbiota with one another and with their dogs. {\em eLife} {\bf 2}.
%
\bibitem{Sul11}
Sul WJ, Cole JR, Jesus ED, Wang Q, Farris RJ, Fish JA et al (2011). Bacterial community comparisons by taxonomy-supervised analysis independent of sequence alignment and clustering. {\em Proc Natl Acad Sci} {\bf 108}: 14637-14642.
%
\bibitem{Tourova03}
Tourova TP (2003). Copy number of ribosomal operons in prokaryotes and its effect on phylogenetic analyses. {\em Microbiology} {\bf 72}: 389-402.
%
\bibitem{Turnbaugh10}
Turnbaugh PJ, Quince C, Faith JJ, McHardy AC, Yatsunenko T, Niazi F et al (2010). Organismal, genetic, and transcriptional variation in the deeply sequenced gut microbiomes of identical twins. {\em Proc Natl Acad Sci}  {\bf 107}: 7503-7508.
%
\bibitem{VandeWalle12}
VandeWalle JL, Goetz GW, Huse SM, Morrison HG, Sogin ML, Hoffmann RG et al (2012). Acinetobacter, Aeromonas and Trichococcus populations dominate the microbial community within urban sewer infrastructure. {\em Environ Microbiol} {\bf 14}: 2538-2552.
%
\bibitem{Youngblut13}
Youngblut ND, Shade A, Read JS, McMahon KD, Whitaker RJ (2013). Lineage-Specific Responses of Microbial Communities to Environmental Change. {\em Appl Environ Microbiol} {\bf 79}: 39-47.
%
\bibitem{Zhang00}
Zhang Z, Schwartz S, Wagner L, Miller W (2000). A greedy algorithm for aligning DNA sequences, {\em J Comput Biol} {\bf 7}: 203-214.
%
\bibitem{Zheng12}
Zheng ZJ, Kramer S, Schmidt B (2012). DySC: software for greedy clustering of 16S rRNA reads. {\em Bioinformatics} {\bf 28}: 2182-2183.
%
\end{thebibliography}
\end{document}